\documentclass[10pt, a4paper]{article}
\usepackage[dvipsnames,table,xcdraw]{xcolor}
\renewcommand{\arraystretch}{1.4}
\usepackage[margin=1in]{geometry} 
\usepackage[utf8]{inputenc} 
\usepackage[T1]{fontenc} 
\usepackage[labelfont=bf, skip=5pt]{caption}
\usepackage{subcaption}
\usepackage{graphicx}
\graphicspath{{images/}}
\usepackage{eso-pic}
\newcommand\LowerRight[1]{\AtPageLowerLeft{%
\put(\LenToUnit{0.75\paperwidth},\LenToUnit{0.03\paperheight}){#1}}}
\usepackage{dblfloatfix}
\usepackage{enumitem} 
\usepackage{amsmath,amssymb,bbm,mathdots} 
\usepackage{mathtools} 
\usepackage{relsize} 
\usepackage{multirow,chngpage} 
\definecolor{mycolor}{HTML}{8f061a}
\definecolor{urlcolor}{HTML}{5e0702}
\definecolor{highlight}{HTML}{1f049a}

\usepackage[misc]{ifsym}
\usepackage{csquotes}
\usepackage[subfigure]{tocloft}
\usepackage{natbib}
\usepackage[colorlinks = true,
            linkcolor = highlight,
            urlcolor  = urlcolor,
            citecolor = mycolor,
            anchorcolor = highlight]{hyperref}
\usepackage{doi}
\usepackage[hang,flushmargin]{footmisc}
\usepackage{varwidth}
\usepackage[most]{tcolorbox}
\tcbuselibrary{skins}
\newtcolorbox{mybox}[2][]{
  enhanced, breakable,
  before skip=7mm,after skip=7mm,
  colback=black!5,colframe=black!50,boxrule=0mm,
  attach boxed title to top left={xshift=1cm,yshift*=1mm-\tcboxedtitleheight},
  varwidth boxed title*=-3cm,
  boxed title style={
    frame code={
      \path[fill=tcbcolback!30!black]
        ([yshift=-1mm,xshift=-1mm]frame.north west)
        arc[start angle=0,end angle=180,radius=1mm]
        ([yshift=-1mm,xshift=1mm]frame.north east)
        arc[start angle=180,end angle=0,radius=1mm];
      \path[left color=tcbcolback!60!black,right color=tcbcolback!60!black,
            middle color=tcbcolback!80!black]
        ([xshift=-2mm]frame.north west) -- ([xshift=2mm]frame.north east)
        [rounded corners=1mm]--
        ([xshift=1mm,yshift=-1mm]frame.north east) -- (frame.south east) --
        (frame.south west) -- ([xshift=-1mm,yshift=-1mm]frame.north west)
        [sharp corners]-- cycle;
    },
    interior engine=empty,
  },
  fonttitle=\bfseries,
  title={#2},#1
}
\newcommand{\parent}[1]{ \left( #1 \right) }

\newcommand{\curly}[1]{ \left\{ #1 \right\} }
\newcommand{\abs}[1]{ \left| #1 \right| }
\DeclareMathOperator*{\argmax}{argmax}
\renewcommand{\o}{\omega}

\renewcommand{\d}{\delta}
\newcommand{\Th}{\mathbf{T_h}}
\newcommand{\bunderline}[1]{\mkern4mu\underline{#1\mkern-0mu}\mkern5mu }
\newcommand{\A}{\mathcal{A}}

\begin{document}

\setcitestyle{round}
\captionsetup{width=.9\linewidth, font={scriptsize}}

\AddToShipoutPictureBG{
  \LowerRight{{\includegraphics[width=3cm,keepaspectratio]{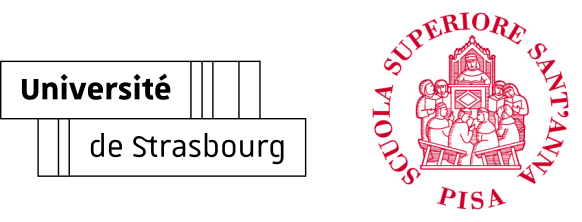}}}
}%

\begin{center}
   \rule{\textwidth}{.2cm}\\ \vspace{.5cm}
   \huge{{Two halves don't make a whole: instability and idleness emerging from the co-evolution of the production and innovation processes}} \\ 
   \vspace{.6cm} \hrule \vspace{.8cm}
    \Large{Patrick Llerena\textsuperscript{1}, Corentin Lobet\textsuperscript{2}, André Lorentz\textsuperscript{1}} \\[10pt]
\end{center} 

 \small
    \noindent
    \textsuperscript{1} Bureau d'Economie Théorique et Appliquée, University of Strasbourg, University of Lorraine, CNRS, 61 Avenue de la Forêt Noire, 67000, Strasbourg, France\\
    \textsuperscript{2} Institute of Economics and L'EMBEDS, Scuola Superiore Sant'Anna, Piazza Martiri della Libertà 33, 56127, Pisa, Italy\\[5pt]
     \Letter $\ $ \href{mailto:pllerena@unistra.fr}{pllerena@unistra.fr}; \href{mailto:corentin.lobet@santannapisa.it}{corentin.lobet@santannapisa.it}; \href{mailto:alorentz@unistra.fr}{alorentz@unistra.fr} 
     \\[0pt]
 \normalsize

\begin{mybox}[colbacktitle=black]{Abstract}
   {\normalsize We propose a disaggregated representation of production through an agent-based fund-flow model (NGR-ADAPT) within which inefficiencies, such as factor idleness and production instability, emerge from endogenous frictions. The model incorporates productivity dynamics (learning and depreciation) and is extended with time-saving process innovations. Specifically, we assume that workers possess inherent creativity that flourishes during idle periods. The firm, rather than laying off idle workers, is assumed to exploit this potential by involving them in the innovation process. Results show that a firm’s organizational and managerial decisions, the temporal structure of the production system, the speed at which workers learn and forget, and the pace of innovation are critical factors influencing production efficiency in both the short and long run. The co-evolution of production and innovation processes emerges in our model through the two-sided effects of idleness: whereas it drives skill decay it is also a condition for creative thinking that can be leveraged for innovation. In doing so, we question the utilization of labour as an adjustment variable in a productive organisation. The paper concludes by discussing potential solutions to this issue and suggesting avenues for future research.}
\end{mybox} 

\hfill
\hfill\\
\hfill


\section{Introduction}

    \textbf{On Idleness} Active idleness\footnote{Active idleness refers to leisure and self-realization activities that, unlike inactivity, contribute to personal flourishing. It is related to the concepts of \textit{schole} in Ancient Greek and \textit{otium} in Latin.} was praised long before \cite{russell1932praise} famously did in a dedicated essay. Ancient Greek and Roman philosophers considered it central to philosophical contemplation, the pursuit of virtue, and happiness. Plato and Aristotle believed it was necessary for contemplation and for achieving the highest form of human flourishing, \textit{eudaimonia} \citep{samaras2017leisure}. Nonetheless, their perspectives differed significantly on whom idleness was meant for. In the \textit{Republic}, Plato argued that the working class (\textit{producers}) should focus on productive activity, enabling the ruling class to engage in idleness. Although he respected work, he believed that each class should remain devoted to its role---workers engaging in contemplation and virtue would cause societal disequilibrium (\textit{injustice}).\footnote{\enquote{that one man should practise one thing only, the thing to which his nature was best adapted [...] justice [is] doing one’s own business [...] there are three distinct classes, any meddling of one with another, or the change of one into another, is the greatest harm to the State, and may be most justly termed evil-doing} \citep{platorepublic}.} Aristotle, in contrast, viewed work not as an end in itself, but as a means to happiness \citep{owens1981aristotle, balme1984attitudes}, a perspective later adopted by thinkers like Bertrand Russell.

    The role of idleness in the workplace has also been a subject of debate. While Seneca emphasized the importance of inactive idleness for rest and relaxation,\footnote{\enquote{our spirits should have time for relaxation: they return from rest better and keener than before. [...] constant hard work breaks the power of spirits. They gain back their strength after taking a break and resting for a while} \citep{senecamind}} the pursuit of profit led factory and mine owners to prioritize productivity---often at the expense of workers who historically worked long hours. This could include women and children \citep{tuttle2010child} who became more easily employable thanks to the Smithian division of labour that gave rise to less strenuous tasks. 
    
    The culture of efficiency reached its zenith in the late 19th and early 20th centuries as Frederick Taylor advocated for the optimization of the factory. Taylor promoted the repetition of efficient moves by a highly specialized and divided workforce, all under the strict monitoring of the chronometer \citep{taylor1911principles}. His ideas are known to have inspired production paradigms such as the Fordist and Toyotist factories and, more generally, lean manufacturing.\footnote{Historical facts and interpretations expressed here are in part based on the historical documentary \textit{Le Temps des ouvriers} by Stan Neumann.}   
    
    In stark contrast, Russell later praised idleness in an essay combining a critique of historical labour conditions (and their unethical foundations) with an argument on the benefits of active idleness and leisure for individuals and society at large. Russell called for the establishment of shorter working hours—roughly equivalent to Keynes' prediction of 15-hour work weeks \citep{keynes1930economic}. Russell observed that, in response to productivity gains and stable demand, firms tended to lay off \textit{unnecessary} workers rather than reducing working hours---a strategy benefiting capital owners, shareholders, and white-collar workers at the expense of the working class. Instead, he suggested that extra productivity should lead to reduced working hours, allowing all individuals to engage in active idleness, which he saw as the key to a fulfilling life.

    In this paper, we propose a different approach to the management of idleness, one more compatible with perpetual profit-seeking. Rather than firing workers—whose efforts made the process efficient—firms could involve them in the innovation process, thereby fostering further efficiency. Assuming active idle time is key to creativity, firms could harness workers' innovative ideas by encouraging a constructive and creative use of their idle time at work.

    \hfill\\
    \textbf{Production and the Firm} In economics, there is a classical divide between market and production, with an increasing gap favouring the market over production. Typically, the theory of production is restricted to discussions surrounding the production function and the theory of the firm as an entity that maximizes profit under constraints, with the production function being one of those constraints. \cite{dreze1985uncertainty} aptly frames this diagnosis: \enquote{The firm fits into general equilibrium theory as a balloon fits into an envelope: flattened out! Try with a blown-up balloon: the envelope may tear, or fly away: at best, it will be hard to seal and impossible to mail… instead, burst the balloon flat, and everything becomes easy.} 

    Since at least the 1930s, particularly with Coase’s seminal work \citep{coase1937nature}, numerous contributions have attempted to conceptualize and model both production and the firm (though many have done so in incomplete or unsatisfactory ways). Among the various attempts to open the black box of the firm and production, two primary trends can be identified in the literature. The first views the firm as an entity that solves inherent issues related to organization, performance, and change within the firm, which only incidentally happens to produce and interact with its environment. The second focuses on the process of production, whose \textit{raison d'être} is the transformation and combination of commodities, labour, and machinery into other commodities, artifacts, or services, which incidentally happen to occur within a firm. 

    More precisely, the first one deals with the organizational aspects of the production–from transactions to learning and routines–the most recent development being the evolutionary and knowledge-based approaches to firms \citep{winter2006toward, nelson2006commentary, dosi2007perspective, marengo2020organizational}. The main perspective is to stress the importance and the role of productive knowledge and learning processes. However, strong critiques have been raised against knowledge-based approaches, particularly as regards their ability to address contemporary economic contexts, such as the rise of disruptive digital technologies or creative behaviour \citep{alvarez2020developing, cohendetetal2024}.

    The second approach focuses on the analysis of production itself, emphasizing the production function as a means to combine production factors. It critiques the simplistic versions of the production function offered by the neoclassical tradition. The Cambridge Controversy, which emerged after WWII, questioned the nature of production factors---particularly capital, whose heterogeneous nature requires an integration into production that transcends mere accumulation \citep{robinson1953production}. This line of thought also raises concerns about the combination of production factors and the impacts on efficiency. For instance, the classical assumption of substitutability of factors is at odds with the empirical evidence \citep{kaldor1957model}. \cite{sraffa1961production} further highlighted the interdependence within and between production processes, stressing the physical constraints that shape production at the firm, sector, and economy-wide levels \citep{pasinetti1973notion}.

    More recently, this literature has evolved into two streams. The first, following \cite{hildenbrand1981short}, critiques short-run production functions and their relevance to representing the technological characteristics of an industry through aggregation.\footnote{\enquote{short-run efficient production functions do not enjoy the well-known properties which are frequently assumed in production theory. For example, constant returns to scale never prevail, the production functions are never homothetic, and the elasticities of substitution are never constant. On the other hand, the competitive factor demand and product supply functions [...] will always have definite comparative static properties which cannot be derived from the standard theory of production.} \citep{hildenbrand1981short}.} The second stream focuses on directly representing production processes as sequences of productive activities and tasks, each characterized by factors such as duration and productivity. Notably \cite{georgescu1970economics} provided a first interesting framework known as the \textit{fund-flow model} of production.

    \hfill\\
    \textbf{The Fund-Flow Approach} Half a century ago, Nicholas Georgescu-Roegen (henceforth NGR) criticized classical quantitative and flow-based\footnote{I.e., relating inputs $\mathbf I$ and outputs $Q$ in quantities, $Q=F(\mathbf I)$, or input rates $\mathbf i=\mathbf I/T$ to output rates $q=Q/T$, $q=f(\mathbf i)$.} production functions for they obscure the underlying process---specifically the role of durations \citep{georgescu1970economics}. This abstraction conceals inefficiencies inherent in the organization of production, that, somewhat independently of factor availability, can disrupt the output rate. In the fund-flow model originally proposed by \cite{georgescu1970economics, georgescu1971entropy} and later studied and refined by \cite{tani1988flows, morroni1992production, piacentini1995time, mir2007funds}, among others, these inefficiencies primarily manifest as funds idleness, i.e., periods of production factors inactivity.

    Funds are defined as factors that remain unaltered by the production process and serve to process the flows, that, in contrast, are altered to create the final product. Funds typically include land, machines, and labour, while flows encompass any input or combination of inputs meant to be transformed---from raw materials to marketable commodities. However, because we relax the strict assumption of fund \textit{sameness}\footnote{The definition and treatment of funds, notably machines and workers, has been criticized by \cite{lager2000production} and \cite{kurz2007fund}. NGR was aware of the limiting nature of the constant productivity assumption but he chose to abstract from it.}---allowing the productivity of workers and machines to change dynamically in response to their service—we redefine funds as production factors that uphold, store, and act upon flows without being physically incorporated into the final good.\footnote{One of the limitations of this definition concerns the production of services: a service can incorporate and alter the workers to a certain extent. More generally, we do not model labour exhaustion. On the other hand, we emphasise that the model proposed in this paper provides a suitable platform for the modelling of the alteration of workers and the initiatives they might undertake to defend their integrity.}
    
    NGR called for the arrangement of production systems \textit{in-line}, i.e., in a modern factory fashion, which seeks to eliminate the idleness of production factors often found in sequential organizations. In sequential systems, idleness arises because subprocesses (also referred to as phases, stages, or tasks) have different durations, causing some phases to \textit{wait for the others}. Production lines are designed to increase the productive capacity of slower phases by allocating relatively more funds to them. By equalizing durations, the process should ideally stabilize and run continuously and efficiently by eliminating idleness. This approach aligns closely with Taylorist views on industrial production. Assuming no additional friction, the original fund-flow model demonstrated the economic superiority of factory-style production processes from an organizational perspective. In stark contrast, we aim, in this paper, to introduce frictions, which, as opposed to the traditional external shocks modelled as if coming from the firm's environment, endogenously originate within the firm.

    \hfill\\
    \textbf{Outline} The paper is composed of two main parts: Sections 2 and 3 c present an agent-based model (ABM) of the fund-flow framework, retaining its core principles while introducing several key departures. We discretize the model, introduce productivity dynamics, and emphasize the principles of factor indivisibility and complementarity, giving rise to the NGR-ADAPT model. We then conduct several experiments to explore the core emerging dynamics of in-line production. In Section 4, we augment the base model with mechanisms that allow for idleness-driven process innovations. These are assumed to be time-saving as they affect the durations of the production phases. We investigate the interplay between idleness-driven innovations and productivity dynamics in determining the process' efficiency, especially in scenarios where workers are assumed to forget as they remain idle for extended periods (skill decay). The results are mixed and the controversial feasibility of this factory paradigm opens the way to new research perspectives. In Section 5, we outline promising directions for further exploration of this topic and propose additional ways in which the NGR-ADAPT model can contribute to research in both microeconomic and macroeconomic theory.


\section{Base Model: NGR-ADAPT}

    The original fund-flow model is a time-continuous representation in which production results from the cumulative service of funds and the processing of flows. NGR highlighted several key characteristics shared by many production systems. First, the production of a commodity often requires the execution of multiple phases that usually differ in skills and duration. Second, he observed that this sequence of subprocesses often leads to the idleness of production factors and he argued that in-line organizations could eliminate such idleness. In this sense, process and organizational innovations could be deemed time-saving \citep{morroni1992production, piacentini1997time, von1995time}. NGR understood that economies of time are central to the development of industries, hence the express representation of durations in his model. For a an in-depth review of the fund-flow model and its extensions refer to \cite{marzetti2013fund}.

    In our implementation, we build on these core principles while also departing from some of NGR's assumptions. First of all, we adopt a discrete dynamical framework that can be simulated with computers. Additionally, we introduce frictions that affect both the productivity and idleness of funds. While NGR acknowledged that the complete elimination of idleness is not always feasible, he assumed that, when it is possible, a flawless process, rid of idleness, would emerge. Production systems, however, are rarely frictionless and the arrangement of phases alone is not sufficient to remove idleness. Specifically, we will consider (i) the existence of a tempo\footnote{It is a fair assumption given that in a production process intermediary products take time to travel across the plant, if not multiple sites. Besides, machines and software might need to be reset, workers might need to move and grab raw materials and tools, etc.}, corresponding to the time unit, that limits the pace of operations, (ii) productivity dynamics for funds (machines and workers), and (iii) the explicit indivisibility of production factors. In these settings, short-term instability and idleness endogenously arise. These frictions can also disrupt NGR's assumption of constant returns to durations: as slowdowns propagate through the system, rather than sparking randomly, the output of $2n$ periods may not necessarily equal twice the output of $n$ periods.\footnote{Note that NGR did not assume constant returns to scale, i.e., to the quantity of employed funds and injected flows.}

    We shall name the proposed model NGR-ADAPT, in honour of NGR, and to emphasize the firm's need to adapt to local slowdowns in the production process, as will become clear. ADAPT stands for \textit{Agent-based DisAggregated Production with Time}.

    \subsection{Sequential Process and Idleness}\label{sec:seq}

        A unit of the final good is produced through the sequential completion of $H$ production phases, altogether an \textit{elementary process}. Assuming that funds operate at their maximum productivity, completing the entire process requires $T = \sum_h T_h$ periods, where each phase $h$ is completed in $T_h < T$ periods. Each phase is executed by a duo of funds $\left(i, j^h\right)$ consisting of a worker $i$ and a machine $j$ of type $h$.

        The quantity of final goods produced after the completion of $T$ periods is normalized to $Q_H=1$. Similarly, intermediate goods $Q_h$ (outflows) are also standardized to 1, and the productivity levels of funds range between 0 and 1. Hence, to produce one unit of $Q_h$ one duo with a combined productivity of 1 must work on task $h$ for $T_h$ periods. 
        
\begin{figure}[ht]
            \centerline{
                \includegraphics[width=1\textwidth]{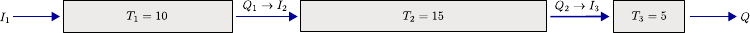}
            }\vspace{.2cm}
            \caption{Sequential Production Process}
            \label{fig:seq}
        \end{figure}

        In phase $h=1$, a duo takes one unit of inflow, $I_1 = 1$, and produces one unit of outflow, $Q_1 = 1$, upon completion of the task. This outflow becomes the inflow for phase 2. All subsequent phases, $h=2,...,H$, thus process an inflow $I_h = Q_{h-1} = 1$ to produce $Q_h = 1$. The outflow of the last phase constitutes the final good. Figure \ref{fig:seq} provides a diagrammatic representation of a sequential process.

        \vspace{.5cm}

        Running this process implies funds idleness as long as there exists a phase $h$ such that $T_h < T_g$ for some $g < h$. The reason is intuitive: if a phase is faster to complete than any of the preceding phases, it needs to wait for the difference in time, $T_g - T_h$. Consider the example in Figure \ref{fig:seq}. If we assume that phase 1 is always supplied with raw materials, then it never becomes idle. Phase 2 is never idle either because the outflows from phase 1 are produced faster ($T_2 > T_1$), resulting in an accumulation of flows between the two phases ($Q_1\equiv I_2$). However, phase 3 remains idle for 10 periods after producing for only 5 periods as it is faster than phase 2. The idleness rate for phase 3 is therefore $(T_2 - T_3) / T_2 = 2/3$. Adding more phases that satisfy $T_h < T_2$ will result in a similar mechanism: these phases will take $T_h$ periods to produce one unit of outflow and will thus be idle $(T_2 - T_h)$ periods every $T_2$ periods. However, if a phase characterized by $T_h \geq T_2$ is added, it will not experience idleness, as it will receive an inflow every $T_2$ periods. In general, under sequential organization and maximum funds' productivity, the idleness rate of any phase $h$ is given by:
        \begin{equation*}
            \mathcal I_h = \frac{T_h^{\max} - T_h}{T_h^{\max}},\quad T_h^{\max} = \max_{g\leq h} \curly{T_g} \label{eq:idle}
        \end{equation*}

        However, this does not hold when the idleness of flows is to be eliminated as well, that is, when the firm wants to prevent the accumulation of inflows (stocks)---a phenomenon that naturally arises in a process with heterogeneous durations \citep{piacentini1995time}. If the firm eliminates stock accumulation, then quicker phases have to wait as well---as opposed to overwork---and the pace is dictated by the slowest phase. For instance, the first phase in our previous example should adapt to the second phase as it is slower and thus should produce one outflow every 15 periods instead of 10. As a result, only the latter is continuously working and all other phases have an idle rate of:
        \begin{equation*}
            \mathcal I_h = \frac{T^{\max} - T_h}{T^{\max}},\quad T^{\max} = \max_h \curly{T_h}
        \end{equation*}

    \subsection{Production Planning}\label{sec:plan}

        \textbf{In-line Organization} The firm can eliminate (or at least mitigate) idleness by organizing the production process in-line. Given the set of integer durations $\Th = \curly{ T_1, \dots, T_H }$, idleness is eliminated by running $C_k = T / \d$ elementary processes every $T$ periods. The resulting quantity of outflows produced, $C_k$, represents the \textit{minimum efficient size} (MES) required to eliminate idleness: the firm cannot achieve maximum efficiency by producing less. The term $\d$ is termed \textit{elementary lag} and corresponds to the inverse frequency (cycle time) at which elementary processes are run. It is equal to the greatest common divisor (GCD) of the durations $\Th$. It follows that the numbers of duos continuously working become $C_h = T_h / \d$. Therefore, a higher MES (lower $\delta$) means that more funds are required to run the process efficiently. 
        
        As a result, the firm produces one outflow of any type $h$ every $\d$ periods and $C_k$ outflows every $T$ periods. In the example of Figure \ref{fig:seq} $T = T_1 + T_2 + T_3 = 30$. The GCD of $(10, 15, 5)$ is $\d = 5$, meaning that by organising in-line, one unit of outflow is produced every 5 periods. This corresponds to $C_k = 6$ elementary processes running every 30 periods, achieved by continuously operating $C_1 = 2$,\ $C_2 = 3$, and $C_3 = 1$ duos in phases 1, 2, and 3, respectively. 
                
        The MES can also be an indicator of planning flexibility. Say for example that we want to target a production of 1.5. It is hard to achieve with an MES of 1 ($\delta=1$) as we could only (efficiently) target multiples of 1, whereas a lower MES enables more accuracy. This is also an important matter when introducing time-saving innovations. Considering an in-line organization, such innovations are not necessarily beneficial as they can impede flexibility and, most importantly, the amount of funds necessary to run efficiently. In the previous example the firm could produce multiples of $C_k=6$ every $T=30$ periods and few funds were required. If we jumped to $T_1=9$ following an innovation on the first phase, then the elementary lag would become $\d=\text{GCD}(9,15,5)=1$ for $T=29$. In this scenario, the MES is therefore 1 and the firm could then efficiently produce multiples of $C_k=29$ every 29 periods, while it would require almost 5 times as many funds to be run efficiently. 

        We generalise the in-line organisation by considering $T_h\in\mathbb R^*_+$ instead of integers. In this case, eliminating idleness completely is rarely feasible\footnote{In order to keep a reasonable MES, the GCD can be generalized to real numbers but it often leads to values close to zero, implying a high MES and the use of a larger number of funds.} so we assume that the firm picks the highest elementary lag $\delta$ in the neighbourhood of durations $\Th$ as it requires fewer funds. Let denote by $\lfloor\cdot\rfloor$ and $\lceil\cdot\rceil$ the floor and ceiling operators, respectively. The firm considers the neighbourhood of integer duration sets:
        \begin{equation}
            \mathbb T_h = \left\{ \mathbf{T_h'}= \{T_1',\dots,T_h,\dots,T_H'\}|_{0\rightarrow 1}  \ :\ T_h'\in\{\lfloor T_h\rfloor,\ \lceil T_h\rceil\}\ \forall h \right\}
        \end{equation}
        
        The operation $|_{0\rightarrow 1}$ replaces all 0s with 1s since the GCD does not exist for sets including 0s. Then the firm picks the highest elementary lag derived in this neighbourhood:
        \begin{equation}
            \d^* = \max \{\boldsymbol{\d'}\} = \max_{\mathbf{T_h'}\in\mathbb T_h} \{\text{GCD}(\mathbf{T_h'})\}
        \end{equation}
        
        The corresponding duration set is denoted by $\mathbf{T_h^*}$. Note that if all durations are real numbers greater than 1 then this approach implies $\d^*\geq2$ since the neighbourhood contains at least one set of even numbers. It follows a natural approximation of the firm's production pace: about $1/\d^*$ units can be produced every period. Letting $T^*=\sum_h T_h^*$---possibly different from $T$---the MES and the number of duos per phase become:
        \begin{equation}
            \mathbf{C_h'} \approx \frac{\mathbf{T_h^*}}{\d^*}, \qquad C_k' \approx \frac{T^*}{\d^*}
        \end{equation}

        \hfill\\
        \textbf{Demand and Parallel Lines} We assume constant demand $\mu_d>0$ as the effect of demand dynamics on production efficiency is not be studied in this paper.
        \begin{equation}
            Y_t^d = \mu_d
        \end{equation}
        
        The firm simply targets a production level of $\mu_d$ per period, or equivalently $T\mu_d$ every $T$ periods. In the ideal case the firm wants to match demand while running an efficient production process, implying to run several lines in parallel\footnote{The lines are nonetheless physically connected, say within a single plant.}. Specifically, the parallelization of the process must satisfy:
        \begin{equation}
            \frac{n}{\d^*} = \mu_d \iff n = \d^*\mu_d
        \end{equation}        
        
        $\mu_d$ may not be an integer while $n$ must be, so we set the number of lines $n^*$ to the nearest greater integer of $n$.        
        \begin{equation}
            n^* = \left\lceil n \right\rceil = \left\lceil \d^*\mu_d \right\rceil 
        \end{equation}
        
        Notice that if $n\in[0, 1]$ then $n^*$ is set to 1, i.e., the firm prefers to produce in excess instead of nothing. 
        
        We finally obtain the number of duos $C_h^*=n^*C_h'$ to allocate to each phase $h$ as well as the size of the process $C_k^*=n^*C_k'$. The firm revises the production process organization every $\tau$ periods. This parameter represents both the firm's reactivity and the timeframe of the process. For example, we set $\tau=50$ for most of the simulations, which would roughly correspond to weekly planning if the time unit is of the order of an hour. Then, a value of $\tau$ of a similar order, e.g. 100 or 200, depict a lower frequency of organisational updates.

    \subsection{Funds Productivity and Management}\label{sec:fund}

        \textbf{Funds Productivity} The working time of funds within a unitary period $(t-1,t]$ is ranging between 0 and 1. If null, the funds either are not allocated to the process or are allocated yet no flow is available for them to work. A working time of 1 on the other hand describes funds that have been working continuously within the period. Finally, the working time can be fractional if an outflow is produced before the end of the period. In this scenario, the duo of funds is both working and idle within the same period. The fractional working time of the duo $(i, j^h)$ is formalized as follows:
        \begin{equation} 
            f_{i/j, h, t} = 
                \begin{dcases}
                    \hfil 1 &\textrm{ if working and } q_{ij,h,t} \leq 1, \\
                    \frac{1 - q_{ij,h,t-1}}{q_{ij,h,t} - q_{ij,h,t-1}} &\textrm{ if working and } q_{ij,h,t} > 1, \\
                    \hfil 0 &\textrm{ otherwise.} 
                \end{dcases}
        \end{equation}        

        Where $q_{ij,h,t}\geq0$ is the current completion rate of the flow processed by the duo (as defined in equation \eqref{eq:prod1}). Working time affects the funds composing the duos in two very opposite way: while workers gain experience through practice machine productivity decays with use. Working time, hence, has a non-trivial effect on the efficiency of the production process. 
        
        Workers' productivity level starts at $a_0\in(0,1]$ and grows as they practice a task, i.e., via \textit{learning-by-doing}.\footnote{Learning-by-doing has been associated with other sources of productivity gains such as organizational and process innovation that do not target the workers routines directly \citep{wright1936factors, newell2013mechanisms}. It has also been modelled through a single proxy embedding many forms of learning \citep{arrow1962economic, solow1997learning}, be it at the firm or at the country level. In contrast, as did \cite{de1957effects} we separate workers' skill improvement (to which we reserve the term learning-by-doing), organizational choices, and process innovations.} 
        For each worker, $a_0$ is set to $a_u$ for phases they are not familiar with and to $a_s\geq a_u$ for the task $h_i^*$ the worker was initially recruited for.\footnote{The parameter $a_s$ can thus be interpreted as a recruitment skill requirement, hence possibly controlled by the firm.} Additionally, a parameter $\gamma_a>0$ controls the learning rate of the worker. The learning curve is assumed to be S-shaped:       
        \begin{equation*}
            a_{i,h,t+1} = 1 - (1 - a_{i,h,t})^{1 + \gamma_a f_{i,h,t+1}}
        \end{equation*}
        
        Several adjustments are made to this equation.\footnote{Note that in order to ensure that the productivity level is not stuck at 1 we artificially add a small value to the 1's making the upper-bound.}
        First, we assume that workers' productivity is floored by $a_u$. Second, we introduce a \textit{forgetting} mechanism: workers who don't practice a task for a long time tend to forget \citep{argote1990persistence, arthur1998factors, shafer2001effects, besanko2010learning}. This mechanism is controlled by a threshold parameter $\theta_a$ such that if $f_{i,h,t+1}>\theta_a$ the worker learns, if $f_{i,h,t+1}<\theta_a$ the worker forgets, and equality is neutral. In doing so, the model reflects the emerging differentiation in skills across the workers.\footnote{In recent years, similar mechanisms of differentiation in workers' skill have been used both in macroeconomic models focusing on the dynamics of the labour market as in \cite{dosi2018causes} and \cite{bordotlorentz2021} and in an evolutionary production model that tackles the adaptation of production to demand cycles \citep{LlerenaLorentzMarengoValente}.} We define: 
        \begin{equation}
            a_{i,h,t+1} = \max\curly{a_u,\ \min\curly{1,\ 
                1.01 - (1.01 - a_{i,h,t})^{1 + \gamma_a (f_{i,h,t+1}-\theta_a)}}}
        \end{equation}        

        In contrast, machines depreciate as they are used. Below is defined the productivity level of a machine $j$ of type $h$ at time $t$, where $\theta_b$ is the depreciation rate and $\mathcal F_{j,h,t}$ is the accumulated working time of the machine since last repair. 
        \begin{equation}
            b_{j, h, t} = e^{-\theta_b \mathcal F_{j,h,t}}
        \end{equation}    

        The firm may decide to repair machines when their productivity level falls below some value $\underline b \in (0,1)$. This operation induces a cost $e_h$ expressed in time---a period during which the machine is not available. The firm waits for the next planning (occurring every $\tau$ periods) to start the maintenance. We assume that more efficient technologies (lower duration $T_h$) are more costly to be repaired.
        \begin{equation}
            e_{h,t} = \frac{\o\tau}{T_{h,t}}
        \end{equation}
        
        For instance, if $\o=10$, then the machines characterized by $T_1=5,\ T_2=10$ and $T_3=20$ will respectively require $2\tau,\ \tau$, and $\tau/2$ periods to be repaired.

        \hfill\\
        \textbf{Allocation} At any instant $t$ the firm is hiring $N$ workers and $J$ machines, split into $J_1...J_h...J_H$ machines of type $1...h...H$. We can represent the pool of funds by a $N\times H$ matrix $\mathbf{L_t^p}$ of workers' productivity levels and by $H$ vectors $\mathbf{K_{h,t}^p}$ of machines' productivity of dimension $J_h$, respectively:\footnote{In practice the firm is unlikely to know the actual productivity of funds at any given time, especially since funds of different nature co-operate. In the simulation model we assume that the firm knows the productivity dynamics of machines and infer workers' productivity. However, this comes down to knowing both perfectly when the dynamics are noise-free, which is assumed throughout this paper.}       
        {\small\begin{equation}
            \mathbf{L_t^p} = 
                \begin{pmatrix}
                    a_{1,1,t} & \dots & a_{1,h,t} & \dots & a_{1,H,t} \\
                    \vdots & \ddots & \vdots & \iddots & \vdots \\
                    a_{i,1,t} & \dots & a_{i,h,t} & \dots & a_{i,H,t} \\
                    \vdots & \iddots & \vdots & \ddots & \vdots \\
                    a_{N,1,t} & \dots & a_{N,h,t} & \dots & a_{N,H,t} 
                \end{pmatrix} \qquad
            \mathbf{K_{h,t}^p} =
                \begin{pmatrix}
                    b_{1,h,t} \\ \vdots \\
                    b_{j,h,t} \\ \vdots \\
                    b_{J_h,h,t} 
                \end{pmatrix}
        \end{equation}}
        
        The targeted number of machines of type $h$ is given by $C_h^* = n^*C_h'$. To reach this level the firm assigns to each phase $h$ the most productive available funds first. Therefore, ideally the first duo assigned to some phase $h$ should be:
        \begin{equation*}
           (i,j^h)^1 = \left( \argmax_i a_{i,h}, \ \ \argmax_j b_{j,h} \right) 
                = \left( \max_i\curly{\mathbf{L_{h,t}^p}}, \ \ \max_j\curly{\mathbf{K_{h,t}^p}} \right)
        \end{equation*}     
        
        Where $\mathbf{L_{h,t}^p}$ is the $h$-th column of $\mathbf{L_t^p}$. If we approach this iteratively, the firm repeats this operation on the remaining available funds until the objective is reached. Formally, letting $M^h$ be the latent amount of duos allocated to phase $h$ and $\A$ be the latent pool of available funds, we obtain the rule:        
        \begin{align*}
            \textrm{Increment } M_h \textrm{ until } 
                \sum_{(i,j^h)\in\A} a_{i,h} b_{j,h} \geq  C_{h,t}^* 
        \end{align*}

        \hfill\\
        \textbf{Adaptation} In this version of the model demand is exogenous and unsatisfied demand accumulates without maximum delay.\footnote{The firm may thus be seen as a monopolist since the clients do not back out.} The firm increases its stock of raw materials $I_1$ by $n^*/\d^*\approx\mu_d$ every period. Importantly, all phases can experience an accumulation of inflows stocks and thus phases that lag behind need to be reinforced with additional workforce. Let $I_{h,t}$ be the stock of inflows in phase $h$ at time $t$. The term $\tau/T_h$ represents the maximum number of flows that can be processed in $\tau$ periods by one duo of funds. Hence the firm may need to allocate $I_h / (\tau/T_h)$ additional duos to phase $h$. This mechanism shall be deemed \textit{reactive adaptation}.  
        
        Besides, the firm may decide to produce at a faster pace in order to offset slowdowns. For this sake, the firm wants to produce $r\geq1$ times its production target. To prevent this from backfiring as a heavy excess production, the firm automatically sets $r=1$ when excess starts being visible (i.e. when $V_{H,t}<0$, see eq. \eqref{eq:V}). While facing delays, however, the firm wants to allocate $rC_h^*$ duos to phase $h$. This mechanism shall be deemed \textit{proactive planning}. 
        
        Accounting for these mechanisms, the allocation of funds becomes: 
        \begin{align}\label{eq:planprodadjusted}
            \textrm{Increment } M_h \textrm{ until } 
                \sum_{(i,j^h)\in\A} a_{i,h} b_{j,h} \geq  \underbrace{rC_{h,t}^*}_{\textbf{proactivity}} + \underbrace{\frac{I_{h,t}}{\tau/T_{h,t}}}_{\textbf{reactivity}}
        \end{align}

        \hfill\\
        \textbf{Initial Pool of Funds} So as to prevent (i) flows from accumulating excessively and (ii) some phases from going short-staffing, the firm seeks to employ a sufficient amount of workers and machines in the first place. We assume that this choice accounts for the optimal number of duos $C_h^*$, augmented by the proactive target factor $r$ and adjusted for the maintenance threshold $\underline b$. The number of workers shall not be too high though. Machines are more numerous due to depreciation, hence backup machines are needed to cope with periods of low machine productivity and of maintenance. We define:      
        \begin{equation}
            J_h = \left\lceil \frac{r C_{h}^*} {\underline b} \right\rceil \qquad\qquad N_h = \left\lceil r C_{h}^* \right\rceil
            \label{eq:inflowsitial}
        \end{equation}

        \hfill\\
        \textbf{Sequential Allocation} Phases that are more likely to accumulate delay may benefit from being assigned funds first so that they can catch up rapidly. We introduce a lexicographic heuristic to define the priority order. The firm first allocates funds to the phase with the highest delay in terms of production delay $V_{h,t}$ (defined in equation \eqref{eq:V}), and proceed likewise for selecting the next phase to prioritize, and so on.\footnotemark If several phases show the same delay, however, the natural order is preserved, i.e., earlier phases are given priority. This choice lies in another important consideration: slowdowns occurring at earlier stages are more harmful for the production system due to the sequential nature of the system---delays in one phase impact all the subsequent phases. 

        \footnotetext{In contrast, the targeted number of duos $C_h^*$ was augmented by the number of inflows stocks $I_h$ in equation \eqref{eq:planprodadjusted}. In fact, both $I_h$ and $V_h$ are reasonable proxies for production delay. However, using either for determining both the priority order and the allocation target $C_h^*$ can lead to bottlenecks. For example, using $I_h$ to set the priority order may cause early phases to capture most or all of the available funds whenever inflows accumulate faster there, which is likely occurring in the early life of the factory or when the firm is proactive ($r>1$). On the other hand, using delay $V_h$ to augment the target $C_h^*$ can cause the later phases to be provided with many funds even when few or no inflows are ready to be processed.}
        
        We can finally formalize allocation. Starting from the highest-priority phase, say $h$, let $\mathbf{\tilde L_{h,t}^p}$ be the $h$-th vector of workers' productivity sorted in descending order and $\mathbf{\tilde K_{h,t}^p}$ be the corresponding vector of machines' productivity. We compute the combined productivity vector for phase $h$:
        \begin{equation}\label{eq:allocate}
            \mathbf{\tilde P_{h,t}} = \mathbf{\tilde L_{h,t}^p} \bunderline\odot \mathbf{\tilde K_{h,t}^p}
        \end{equation}        
        
        Where $\bunderline\odot$ is the Hadamard product applied after truncation of the longest vector such that both have the same dimension. Let $\mathbf{\tilde P_{h,t}^c}$ be the cumulative sum of the productivity vector $\mathbf{\tilde P_{h,t}}$. $M_h$ is then equal to the index of the first element of $\mathbf{\tilde P_{h,t}^c}$ that is greater or equal than the threshold defined in \eqref{eq:planprodadjusted}. If not possible, $M_h$ is equal to the size of the vector, i.e., the number of available funds duos. Then, the productivity vectors are recomputed after dropping the previously allocated workers and the exact same process is applied to the next phases.

    \subsection{Production Mechanics} \label{sec:prod}

        \textbf{Indivisibility} The low level of abstraction in the NGR-ADAPT model stems from the willingness to allow instabilities to emerge endogenously in a disaggregated production model. A key feature that distinguishes our model from NGR’s original framework, and most production models in general, is the explicit representation of indivisibility of production factors. Although they were recognized and discussed by NGR and later contributors to the fund-flow model (see notably \cite{morroni1992production}), indivisibility were not so central in the analysis. Indeed, under the hypothesis that production lines run flawlessly, the express representation of the identity of factors does not add depth to the model, productivity is \textit{reductionist}, and the aggregation of production is thus tractable through the \textit{sum of parts}. 
        
        On the contrary, in our framework aggregation is tedious as duos operate at different paces. There, indivisibility truly matter and the term \textit{two halves don't make a whole} takes full meaning. Notably, the following dimensions of indivisibility are embedded in the NGR-ADAPT model: 
        \begin{enumerate}
            \item Independently processed fractions of the final good usually cannot be added together and sold as one unit, i.e., the output results from the completion of an elementary process involving all phases. For instance, a car worth \$20k is a whole and no customer is willing to acquire two half-completed cars nor do they want to purchase \$20k worth of random components of the car.
            \item This often hold at the phase-level: two bottom halves of a car's door cannot be assembled into one full door ready to be appended to the car.  
            \item Funds (workers and machines) use their productive power for at most one elementary process at any given time.\footnote{This property of the model is sometimes restrictive. While it is common that workers are individually paired with one workstation, computer, or vehicle for example, it can also be true that a duo contributes to several outputs at once (e.g., cheese production) or that a machine is shared (e.g., an oven). Heterogeneity in the nature and interaction of funds is left for future work.} 
        \end{enumerate}

        These indivisibility properties imply \textit{strong complementarity} between production factors. Although the economic literature recognizes the complementarity of production factors, the mechanisms underlying complementarity are typically obscured. Here, simply having sufficient hypothetical production capacity from machines and workers is not enough. The mapping from input factors to output products is non-monotonous and, in fact, it is not even unique. Relatedly, returns to scale and to duration are variable over time due to inefficiencies. 
        
        \hfill\\
        \textbf{Production} Consider the vector of duos $\mathbf{{F_{h,t}^*}}$ allocated to phase $h$ and their productivity vector $\mathbf{{P_{h,t}^*}}$. Both are sorted in descending order just after allocation so as to guarantee that inflows are passed on to the most productive duos first.         
        \begin{equation}
            \mathbf{{F_{h,t}^*}} = 
                \begin{pmatrix}
                    \left( i, j^h \right)^1_\tau \\
                    \vdots \\
                    \left( i, j^h \right)^{n}_\tau \\
                    \vdots \\
                    \left( i, j^h \right)^{N_h}_\tau
                \end{pmatrix} \qquad
            \mathbf{{P_{h,t}^*}} = 
                \begin{pmatrix}
                    p_{h,t}^1 \\
                    \vdots\\
                    p_{h,t}^n \\
                    \vdots \\
                    p_{h,t}^{N_h}
                \end{pmatrix} = 
                \begin{pmatrix}
                     a^1_{i,h,t} \cdot b^1_{j,h,t}  \\
                    \vdots\\
                     a^n_{i,h,t} \cdot b^n_{j,h,t}  \\
                    \vdots \\
                     a^{N_h}_{i,h,t} \cdot b^{N_h}_{j,h,t} 
                \end{pmatrix}
        \end{equation}        
        
        To ensure the independence of factors, production is represented by a latent vector $\mathbf{\tilde Q_{h,t}}$. In any period $t$ the latent production of an activated duo corresponds to the cumulative completion of the flow. Letting $\tilde I_{h,t}^n$ be the latent inflows stock, we write:
        \begin{equation} \label{eq:prod1}
            \mathbf{\tilde Q_{h,t}} =
                \begin{pmatrix}
                    q_{h,t}^1 \\
                    \vdots\\
                    q_{h,t}^n \\
                    \vdots \\
                    q_{h,t}^{N_h}
                \end{pmatrix} \quad 
            \textrm{where} \quad q_{h,t}^n = 
                \begin{dcases}
                    q_{h,t-1}^n + \frac{p_{h,t}^n}{T_h} & \text{ if } q_{h,t-1}^n  \in (0,1), \\
                    \hfil \frac{p_{h,t}^n}{T_h} & \text{ if } q_{h,t-1}^n \notin (0,1) \text{ and } \tilde I_{h,t} \geq 1, \\
                    \hfil 0 & \text{otherwise.}
                \end{dcases}
        \end{equation}    
        
        Let us vectorize equation \eqref{eq:prod1}. Define a vector of latent inflows stock $\mathbf{\tilde I_{h,t}}$ of dimension $N_h$ corresponding to inflows availability: $\tilde I_{h,t} \geq 1$. It is constructed such that more productive duos are prioritized, upon availability. Because 1s are set for available funds only, $\mathbf{\tilde I_{h,t}}$ represents the whole condition of the second row in equation \eqref{eq:prod1}: $q_{h,t-1}^n \notin (0,1) \text{ and } \tilde I_{h,t} \geq 1$. We also consider the binary vector $\mathbf{X_{h,t}}$ of dimension $N_h$ whose entries take on the value 1 if $q_{h,t-1}^n \in (0,1)$ and 0 otherwise. We can write:
        \begin{equation}\label{eq:prod}
            \mathbf{\tilde Q_{h,t}} = 
                \left[ \mathbf{X_{h,t}} \odot ( \mathbf{\tilde Q_{h,t-1}} +  \mathbf{{P_{h,t}^*}} / T_h ) \right] + 
                \left[ \mathbf{\tilde I_{h,t}} \odot \left( \mathbf{{P_{h,t}^*}} / T_h \right) \right]
        \end{equation}   
        
        Where $\odot$ is the Hadamard product. When a duo reaches a latent production of 1 or greater, it means that the outflow is available to be sold or to be used as an inflow in the next phase. The phase-level production function is the sum of duos that have reached a latent production of 1 or greater:         
        \begin{equation} \label{eq:prod_vec}
            Q_{h, t} = \sum_{n=1}^{N_h} \mathlarger{\mathlarger{\mathbbm 1}} \left( q_{h,t}^n \geq 1 \right)
        \end{equation}

        \hfill\\
        \textbf{Stocks of Flows} The latent stock of inflows $\tilde I_{h,t}$ is decremented when inflows are loaded (second condition in equation \eqref{eq:prod1}). At the beginning of each period the new stocks of inflows become:        
        \begin{equation*} 
            I_{h,t} =  \begin{dcases}
                I_{1,t-1} + \mu_d - \sum{}_* \mathbf{\tilde I_{1,t-1}} & \text{ if } h=1 \\ 
                I_{h,t-1} + Q_{h-1, t-1} - \sum{}_* \mathbf{\tilde I_{h,t-1}} & \text{ if } h>1  
            \end{dcases}
        \end{equation*}
        
        Where $\sum{}_*$ stands for the sum of vector entries. In order to reach the proactive production objective the firm needs to inject $r\mu_d$ raw materials every period. Because the quantities of flows need to be integers we define a latent supplement variable:
        \begin{align}
            \tilde R_t = \tilde R_{t-1} - \lfloor \tilde R_{t-1} \rfloor + (r-1)\mu_d,\qquad 
                R_t = \lfloor \tilde R_t \rfloor
        \end{align}    
        
        Each period the firm increases $\tilde R_t$ by an amount $(r-1)\mu_d$ of supplementary inputs. When this variable becomes greater than 1, the integer part is stored in $R_t = \lfloor\tilde R_t\rfloor$ and the latent variable $\tilde R_t$ is decremented by the same amount. The complete stock dynamics are then:        
        \begin{equation} \label{eq:inflows}
            I_{h,t} =  \begin{dcases}
                I_{1,t-1} + (Q^*_t + R_t) - \sum \mathbf{\tilde I_{1,t-1}} & \text{ if } h=1 \ \& \ V_{H,t} \geq 0 \\
                I_{1,t-1} - \sum \mathbf{\tilde I_{1,t-1}} & \text{ if } h=1 \ \& \ V_{H,t} < 0 \\
                I_{h,t-1} + Q_{h-1, t-1} - \sum \mathbf{\tilde I_{h,t-1}} & \text{ if } h>1 
            \end{dcases}
        \end{equation}   
        
        Where the condition $V_{H,t}<0$ (defined in equation \eqref{eq:V}) corresponds to a regime of excess production, in response to which the firm stops injecting raw materials to prevent stocks from accumulating in large quantities.

\section{Emerging Inefficiencies: Baseline Experiments}

    In this first set of simulation experiments, we explore the system dynamics for different configurations of managerial reactivity and proactivity, under various temporal structures, as well as for different degrees of learning and forgetting. In doing so, we first focus on the emergence of inefficiencies, namely production slowdowns and workers' idleness, and the ways to cope with these inefficiencies. 
    
    \subsection{Performance Evaluation} \label{sec:perf}

        The planning and adaptation mechanisms are meant to let the bounded-rational firm demonstrate a decent degree of intelligence with the ultimate goal of converging to a stable and efficient production regime. We thus need to assess the stability and the efficiency of the process. We do so by considering the firm's ability to satisfy demand, on the one hand, and the idleness of workers, on the other. Ideal conditions for a firm are to \textit{quickly converge} to a regime of \textit{quick demand satisfaction with low variability} in production delays and excess while \textit{maintaining workers' idleness low} in the long run (as labour bears a cost for the firm, even though we abstract from it).

        \hfill\\
        \textbf{Demand Satisfaction} First of all, we measure the firm's ability to fulfil demand through the difference between production and demand, denoted $D_h$. We define the accumulated production differential as the sum of these differences and denote it by $V_h$. Specifically, for phase $H$ (the final good), a positive value indicates a lagging production while a positive value corresponds to excess production. 
        \begin{equation}
            D_{h,t} = \mu_d - Q_{h,t}, \qquad
            V_{h,t} = \sum_{z=1}^t D_{h,z} \label{eq:V}
        \end{equation}

        \hfill\\
        \textbf{Idleness} The idleness of a fund is given by $t^-_{i/j, t} = 1-f_{i/j,h,t}$. It is ranging between 0 and 1 if the fund is allocated and null otherwise. The overall idle rate of workers is:
        \begin{equation}
            \text{IRW}_t = \frac{1}{N}\sum_i t^-_{i, t}
        \end{equation}    
        
        We distinguish between intentional and unintentional idleness. Let denote by $\bar\A_t$ the set of allocated duos (as opposed to the set of available duos $\A$). Its cardinality is the number of active workers or machines, i.e., $|\bar\A_t|=\sum_i \mathlarger{\mathbbm 1}(i \text{ is allocated at time } t)$. The overall intentional idle rate of workers is the proportion of non-allocated workers:
        \begin{equation}
            \text{IRW}_t^i = 1 - \frac{|\bar\A_t|}{N}
        \end{equation}
        
        The overall unintentional idle rate averages the idleness of allocated funds:
        \begin{equation}
            \text{IR}_t^u = \frac{\sum_{i\in\bar\A_t} t^-_{i, t}}{|\bar\A_t|}
        \end{equation}

    \subsection{Model Configuration}

        We perform \textit{ceteris paribus} analyses parameters by fixing parameters values unless they are part of the experiment under study. The default environment is parametrized as follows:
        
        \begin{itemize}[partopsep=.8ex, itemsep=0pt]
            \item Simulations are run for 50,000 periods. If 50 periods represent about a week, then this corresponds to about 20 years.
            \item Production involves $H=5$ production phases with equal duration $T_h=6,\ \forall h$, hence $T=30$.\footnote{In this scenario durations are integers and $\delta=6$.}
            \item Demand is fixed to $\mu_d=1$ per period.
            \item Planning and allocation are operated every $\tau=50$ periods.
            \item The initial productivity levels are $a_u=0.2,\ a_s=1$, i.e. the firm recruits specialized workers.
            \item The learning-and-forgetting rate is set to $\gamma_a=0.001$ and the forgetting threshold is set to $\theta_a=0.2$.\footnote{Notice that it implies that workers should spend at least 20\% of their time working on a task to prevent forgetting. So, unless they distribute their time equitably between the five tasks, they will tend forget at least one skill.}
            \item The depreciation rate is set to $\theta_b=0.0002$.
            \item Machine maintenance is triggered when the productivity of a machine goes under $\underline b=.8$ and the maintenance cost parameter is set to $\omega=10$.
            \item The proactivity parameter is set to $r=1.5$.            
        \end{itemize}

        Variables and parameters are summarized in table \ref{table} in Appendix \ref{sec:appendix}.

        Figure \ref{fig:proddyn} illustrates the hypothetical productivity dynamics of funds in the baseline. Assuming a constant working time, it shows how the productivity of a worker is impacted by the working time relative to the forgetting threshold. Productivity declines or remains minimum if the worker spends less than 20\% of their time practising the task. Otherwise, it either remains steady (exactly 20\%) or it improves or remains maximum (more than 20\%). Moreover, the time required to mastering a task is of the order of thousands of hours for workers consistently practising---in line with both empirical findings \citep{newell2013mechanisms} and the popular 10,000-hour rule \citep{simon1988skill, gladwell2008outliers}.

        \begin{figure}[t]
            \centering
            \includegraphics[width=\linewidth]{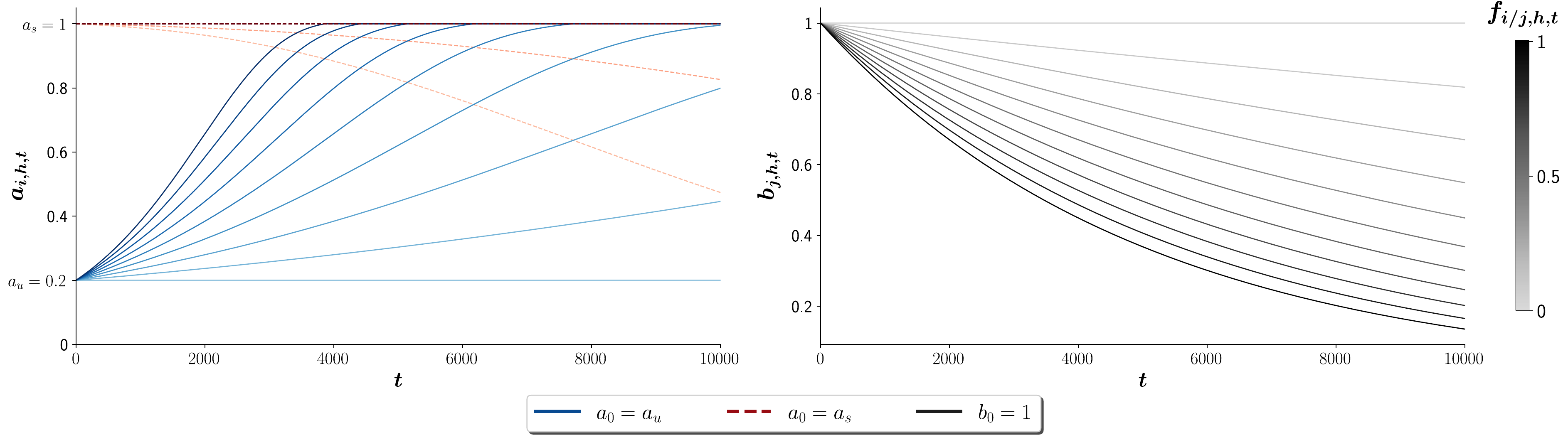}
            \caption{Productivity Dynamics for Baseline Calibration. The left graph shows the productivity of a worker who would spend some time $f_{i,h,t}$ working on task $h$ every period $t$. The color and style distinguishes between a skilled and an unskilled worker for the task, and shade is a cue for the working time. The right graph shows the depreciation of a machine and follows the same logic. Maintenance is omitted.}
            \label{fig:proddyn}
        \end{figure}

    \subsection{Experiment 1: Managerial Choices} \label{sec:exp1}

        Several parameters can be deemed adjustable by the firm. Three, in particular, hold special significance: the periodicity $\tau$ of organizational updates, the maintenance threshold $\underline{b}$, and the proactivity parameter $r$. We let $\tau\in\{10, 50, 1000\}$, $\underline b\in\{0.2, 0.5, 0.65, 0.8, 0.95\}$, and $r\in\{1, 1.25, 1.5, 1.75, 2\}$.
        
\begin{figure}[h]
            \centering
            \includegraphics[width=\textwidth]{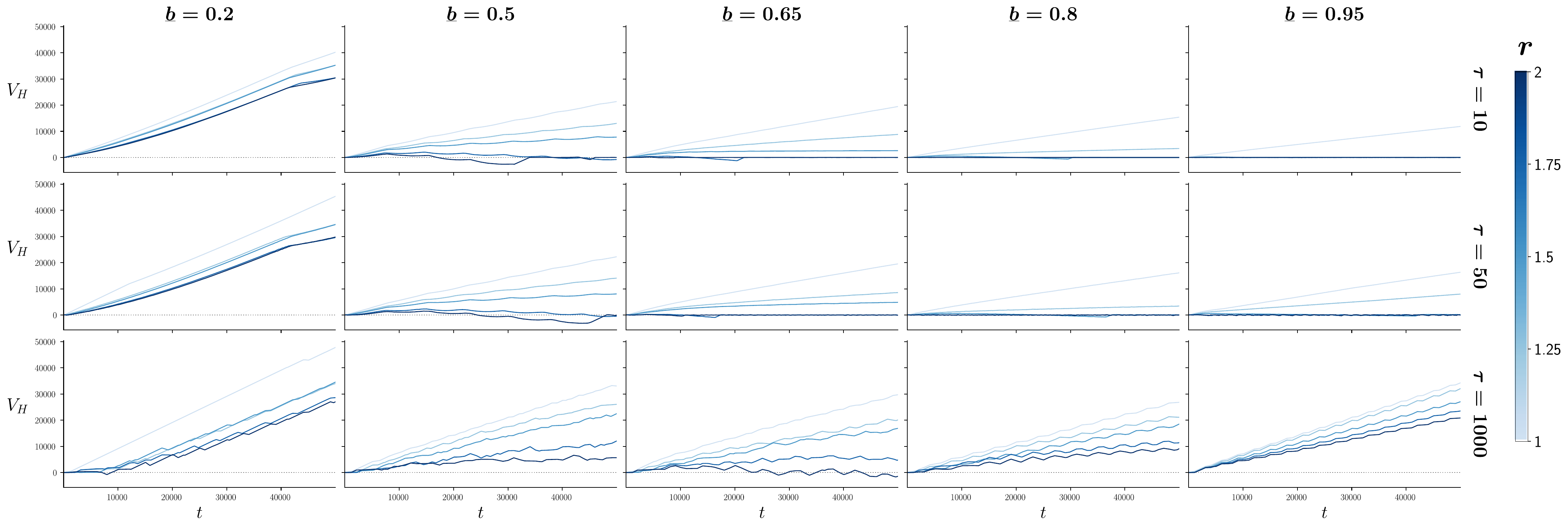}
            \caption{Production Dynamics against Managerial Choices. Is depicted the evolution over time of the variable $V_H$, smoothed over 100 periods (moving average). Rows let $\underline b$ vary, columns let $\tau vary$, the shades let $r$ vary.}
            \label{fig:managVH}
        \end{figure}

        Figure \ref{fig:managVH} presents the evolution of the accumulated production differential indicator $V_{H,t}$ across these 75 configurations. The first clear observation is the impact of proactivity: higher values of $r$ lead to less accumulated delay and a faster convergence to a stable, low-volatility production regime. Second, more frequent organizational updates, reflected by lower values of $\tau$, are also beneficial. Additionally, more frequent maintenance proves advantageous. However, this effect appears to hold only up to a certain point. Specifically, we observe increased delayed demand when both $\tau$ and $\underline{b}$ are high, potentially due to long periods wherein most or all machines undergo maintenance.

\begin{figure}[t]
            \centering
            \includegraphics[width=\textwidth]{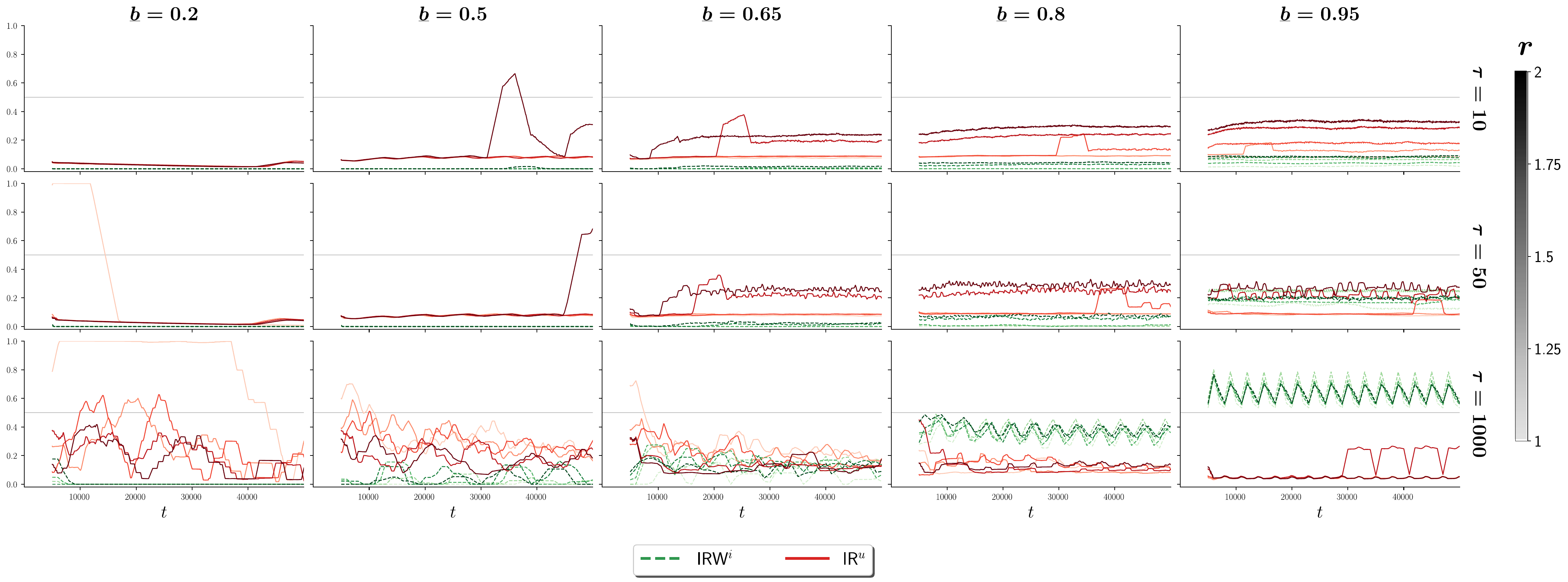}
            \caption{Idleness against Managerial Choices. Is depicted the evolution over time of the variables IRW$^i$ and IR$^u$, smoothed over 5,000 periods (moving average), and differentiated by colours and line style. Rows let $\underline b$ vary, columns let $\tau$ vary, and the shades let $r$ vary.}
            \label{fig:managIR}
        \end{figure}

        The evolution of average idleness is depicted in Figure \ref{fig:managIR}. More frequent maintenance generally leads to greater intentional idleness as it is more likely that large batches of machines are being repaired in the meantime, preventing the allocation of all workers. The association is weak when it comes unintentional idleness, as expected given that the maintenance parameter only affects the allocation of funds. 

        The frequency of organizational adjustments has a pronounced effect on the production dynamics. Typically, higher values of $\tau$ (less frequent adjustments) lead to more intentional idleness and less unintentional idleness. This trend likely interacts with the maintenance threshold $\underline{b}$. Because the firm repairs machines every $\tau$ periods, a higher $\tau$ results in larger batches of machines requiring repair. With fewer operable machines, fewer workers can be allocated, thus increasing the rate of intentional idleness. 

        For moderate values of $\tau$ (first two rows in Figure \ref{fig:managIR}), a more proactive behaviour (higher $r$) is associated with increased unintentional idleness. This is because, once stable, production oscillates between regimes of delayed demand and excess production. The latter are more important when $r$ is high. Since inputs stop flowing in these periods of excess more funds are affected. Proactivity is thus to be tempered: although a larger pool of funds helps to accelerate convergence to stable production dynamics it also induces idleness in the long run. 

    \subsection{Experiment 2: Temporal Structures} \label{sec:exp2}

\begin{figure}[!h]
            \centering
            \includegraphics[width=.9\textwidth]{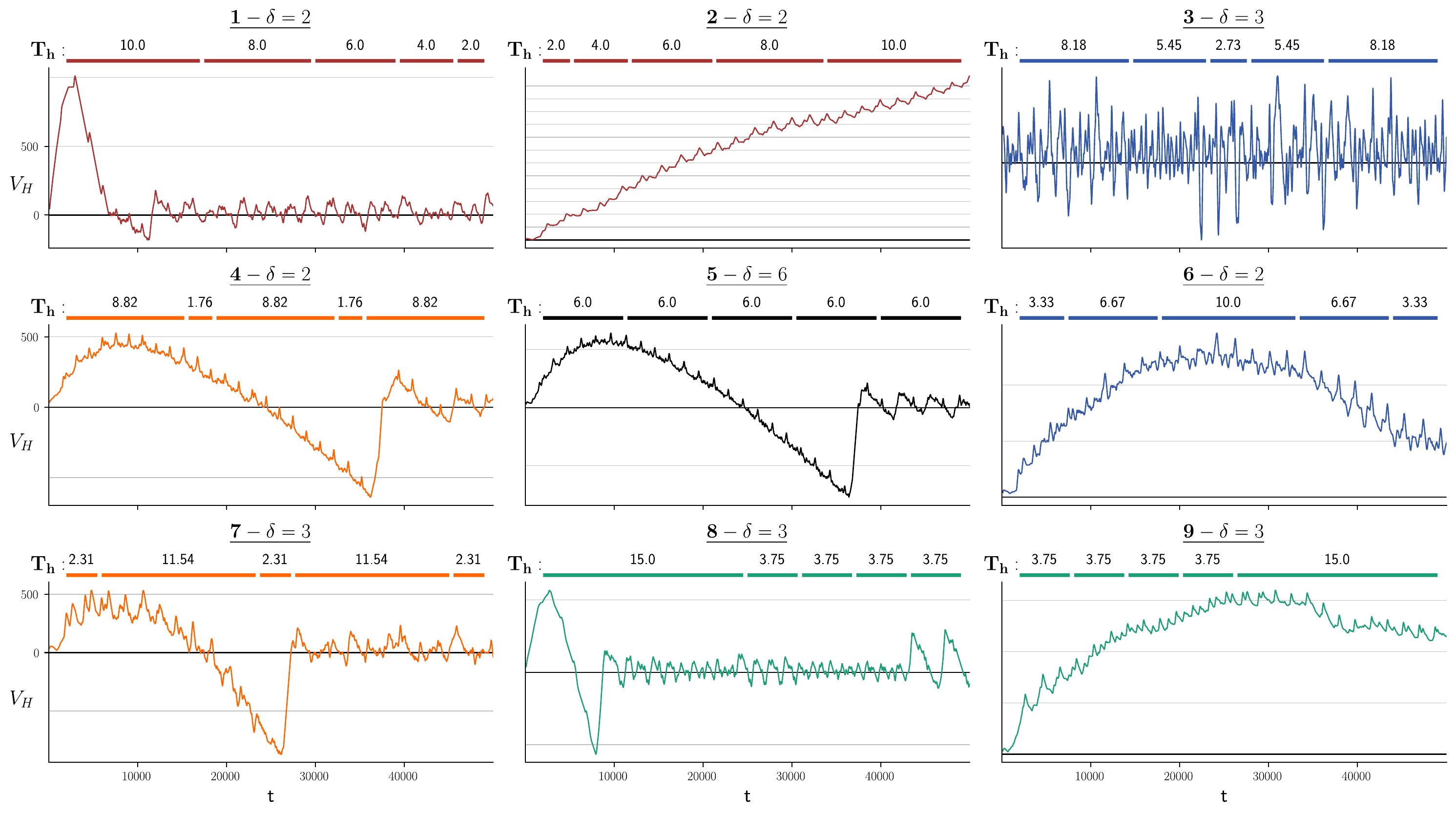}
            \caption{Production Dynamics against Temporal Structures for $T=30$. Is depicted the evolution over time of the variable $V_H$, smoothed over 100 periods (moving average). A black horizontal line is set at 0 and grey lines are repeated every $\pm$500 and serve as cues for comparing the magnitude across plots. The elementary lag $\delta$ is displayed in the subplots' title and the temporal structures $\Th$ are shown under titles. Colors are used to highlight symmetrical structures.}
            \label{fig:timeS30}
        \end{figure}

        The baseline temporal structure, i.e., the arrangement of durations, is uniform: $T_h = 6 \ \forall h$. We now consider eight alternative structures---each of which have the same total duration $T = 30$, grouped into four pairs of symmetrical arrangements. These structures are depicted at the top of each subplot in Figure \ref{fig:timeS30}. The graphs display the evolution of the accumulated production differential $V_H$.
        
        The pair that yields results most similar to the baseline structure (5), which becomes efficient, is the (4-7) pair. All the other pairs exhibit more divergent dynamics. In the peculiar case of the (4-7) pair, a possible determinant of the difference in outcomes is the elementary lag $\d$---structure (4) has a shorter elementary lag, implying fewer funds compared to structure (7). 
        
\begin{figure}[!h]
            \centering
            \includegraphics[width=.9\textwidth]{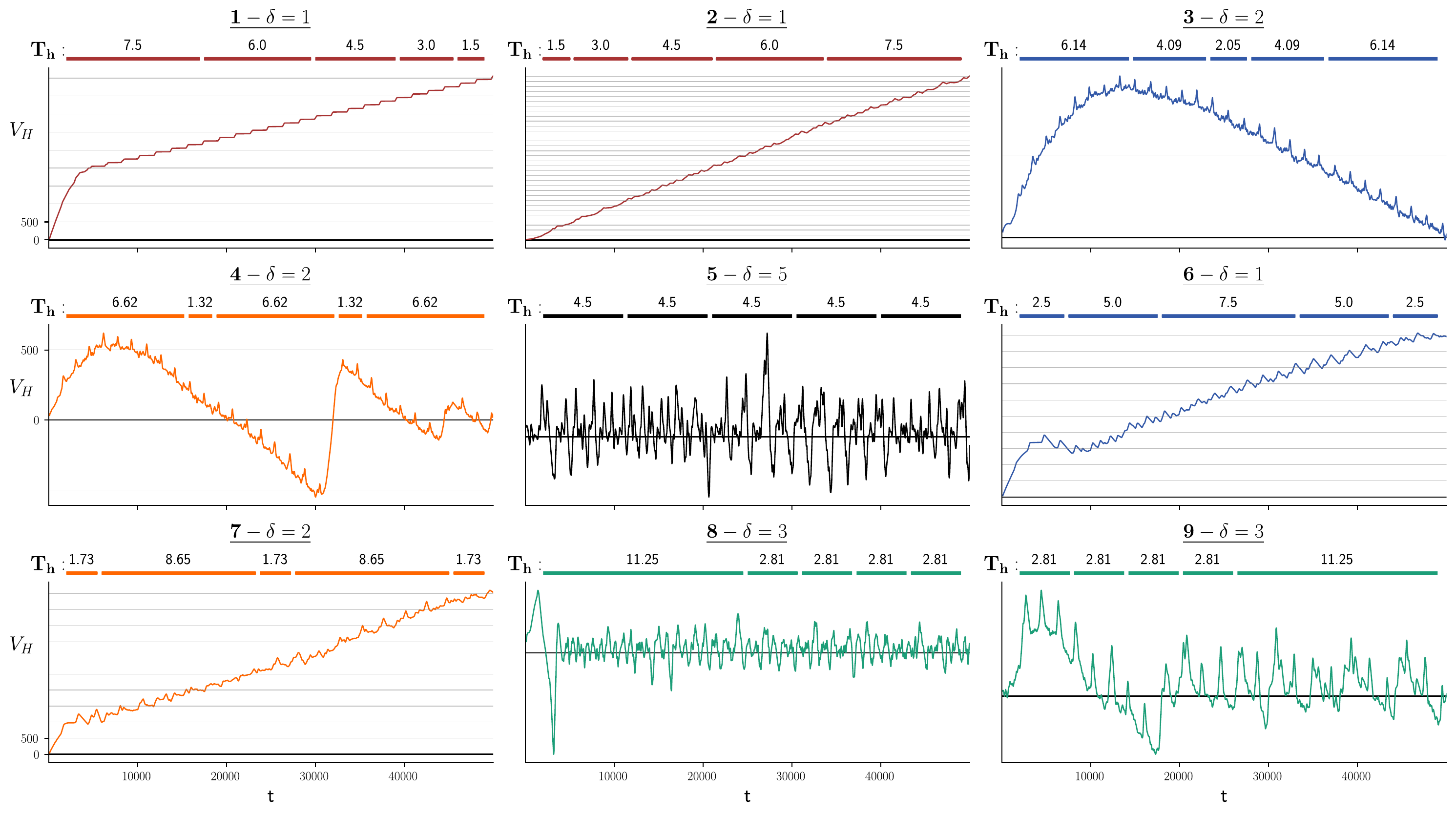}
            \caption{Production Dynamics against Temporal Structures for $T=22.5$. See the caption of Figure \ref{fig:timeS30}.}
            \label{fig:timeS22}
        \end{figure}

        This hypothesis is supported by analysing these same structures with a reduced total duration, $T = 22.5$, as shown in Figure \ref{fig:timeS22}. One might expect faster stabilization in these settings, as seen in the baseline structure (5), but the effects are mixed. While within pairs the winning structure is still the same, not all pairs are affected the same. Notably, pairs (1-2) and (3-6) are worse off, whereas pair (8-9) significantly improved. In the meantime, $\delta$ has decreased for the first two pairs but increases for the last one. Furthermore, in the (4-7) pair, only structure (7) experiences a strong decline in performance, associated with reduced $\delta$, while structure (4) remains relatively unaffected in both regards. Finally, although the baseline structure (5) shows less $\delta$, it demonstrates faster (instant) convergence. However, the proportional decrease in $\delta$ is smaller than in previous cases, which may indicate that the negative effect of reduced $\delta$ diminishes as the initial $\delta$ level increases. We conclude that a lower MES could be beneficial to the system dynamics even though the scenarios wherein the MES is larger have the necessary amount of funds to run efficiently (in theory). 
        
        A deeper investigation into these discrepancies between temporal structures could be a promising direction for future research.

    \subsection{Experiment 3: Learning} \label{sec:exp3}

\begin{figure}[ht]
            \centering
            \includegraphics[width=\textwidth]{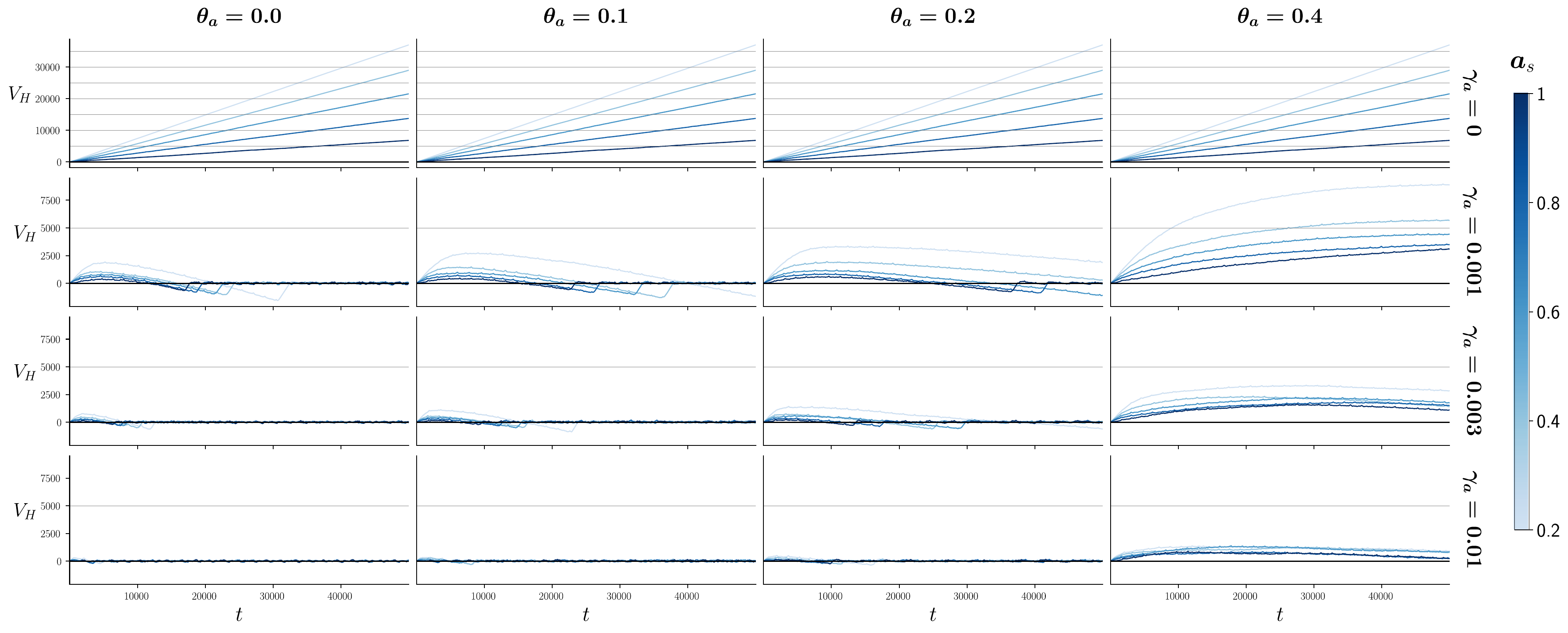}
            \caption{Production Dynamics against Workers' Productivity. The variable $V_H$ smoothed over 100 periods is represented. The parameters $\gamma_a,\ \theta_a$ and $a_s$ are differentiated in rows, columns, and shade, respectively. The y-axis range is common only for the last three rows ($\gamma_a>0$). Horizontal grey lines are drawn every 500 to highlight the magnitudes.} 
            \label{fig:learnVH}
        \end{figure}

\begin{figure}[ht]
            \centering
            \includegraphics[width=\textwidth]{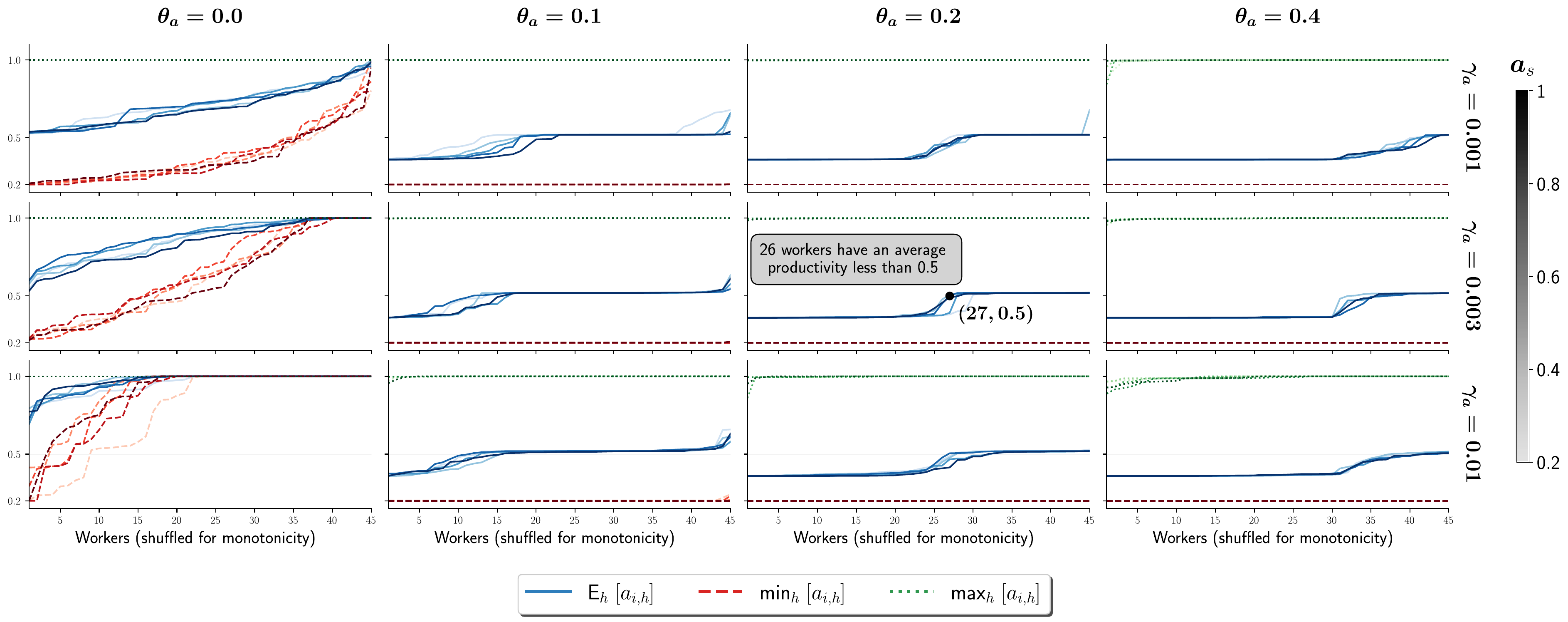}
            \caption{Workers' Specialization. At the last period we compute the mean, minimum, and maximum of each worker's productivity levels. The parameters $\gamma_a,\ \theta_a$ and $a_s$ are differentiated in rows, columns, and shade, respectively. The resulting series (along workers) are independently sorted and graphed. The graphed series inform on the distribution of productivity levels across workers. An example of the adequate interpretation is given in the subplot for ($\theta_a=0.2, \gamma_a=0.001$). \textit{Caution: the x-axis is not representing time.}}
            \label{fig:learnPtf}
        \end{figure}

\begin{figure}[ht]
            \centering
            \includegraphics[width=\textwidth]{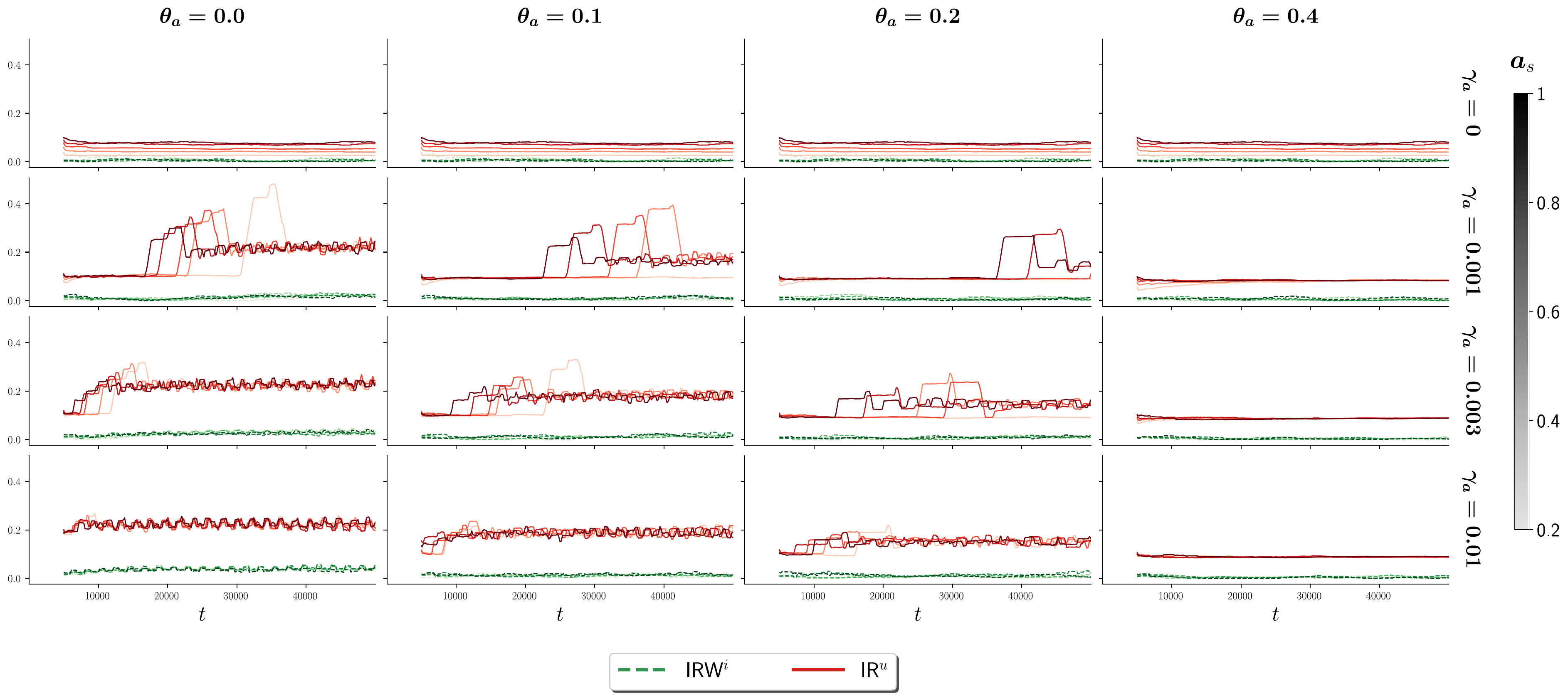}
            \caption{Idleness Rates against Workers' Productivity. The variables IRW$^i$ and IR$^u$ smoothed over 5,000 periods are represented. The parameters $\gamma_a,\ \theta_a$ and $a_s$ are differentiated in rows, columns, and shade, respectively.}
            \label{fig:learnIR}
        \end{figure}

        Workers can learn by doing but also forget if they stop practising a task. To observe the significance of learning and forgetting dynamics we simulate the model in diverse settings by varying the parameters $\theta_a\in\{0, 0.1, 0.2, 0.4\}$, $\gamma_a\in\{0, 0.001, 0.003, 0.01\}$, and $a_s\in\{a_u=0.2,0.4,0.6,0.8,1\}$. The results for the accumulated production differential are depicted in Figure \ref{fig:learnVH} and idleness dynamics are presented in \ref{fig:learnIR}. Moreover, Figure \ref{fig:learnPtf} summarizes the distribution of skills as of the end of the simulation.

        All three parameters have a seemingly important effect on the production dynamics. The least of the three is from the hiring productivity $a_s$. The higher it is the lesser the accumulated delay and the faster the convergence to a stable regime. Starting with more skilled workers indeed places the firm ahead of the game from the beginning. A higher forgetting threshold $\theta_a$ delays the convergence to a stable regime and the learning-and-forgetting rate $\gamma_a$ has a striking impact on the firm's ability to consistently fulfil demand. Interestingly, the positive effect persists for $a_s=1$, shedding light on the importance of skill retention and the acquisition of secondary skills helpful for temporary reinforcements. 

        We can learn more on this by looking at Figure \ref{fig:learnPtf}. We omitted the row for $\gamma_a=0$ as the outcomes are trivial.\footnote{Specifically, the average workers' productivity level would be equal to $(4a_u+a_s)/5$. Depending on $a_s$ this would range between 0.2 and 0.36.} First of all, the hiring productivity has little impact on the final productivity distributions, i.e., despite the accumulated delay is yet to be caught up after the convergence in the distribution of workers' skills. Second, a higher learning rate, as expected, results in more workers having secondary skills. However, this effect is quickly dampened as the forgetting threshold increases. In fact, we can safely state that the effect of the forgetting threshold is predominant. In the extreme case ($\theta_a=0.4$) less than a third of the workers have managed to gain productivity in secondary tasks. 
        
        What seems to emerge from this analysis is that the reversal of $V_H$, which is preceding the convergence to a stable regime, may require workers with a larger skill diversity. Workers are not systematically allocated to the task they have been hired for as confirmed by the persistence of skill diversity in low forgetting scenarios. Workers can be allocated elsewhere to reinforce a phase that has accumulated delay. To be effective, the reinforcement must ideally comprise extra workers who are skilled enough to process the flows quickly. Furthermore, the more productive the reinforcements, the fewer of them will be required in the meantime, such that the other phases do not go short-staffing. This outcome suggests that organisations encouraging skill specialization---at the heart of the division-of-labour paradigms prevailing in Fordist factories--- may not be the most adequate in the long run.

        The average intentional idleness remains fairly low in all configurations so we focus on unintentional idleness. First, the hiring productivity $a_s$ has little effect on long-term idleness rates when the learning rate $\gamma_a$ is non-null. Second, stronger forgetting $\theta_a$ entails less idleness as it impedes the productivity dynamics of workers. Finally, a higher learning-and-forgetting rate seems to induce slightly more idleness. 
        
        Besides, a common pattern is the synchronicity of the large spike in unintentional idleness with the stabilization of the production differential. More precisely, it is contemporaneous with the period of production excess that precedes the stable regime. For some time, the firm needs to slow down production by stopping to feed the process with raw materials $I_1$. Incidentally, the allocated funds face a period of calm and become unintentionally idle. Although not visible in Figure \ref{fig:learnIR} due to smoothing, the idle rate can reach high levels, including 100\%, depending on the extent of production excess.

\section{Productive Idleness through Workers' Creativity}

        Once stabilized, the production process tends to experience higher levels of idleness. A possible way to respond is to rationalize production by reducing the amount of funds employed.  
        Away from situations of economic downturns, a firm may opt to lay off workers even though production objectives are consistently achieved. In line with the ideas of Smith, Taylor, and NGR some dismissals might indeed be an effective solution to eliminate intentional idleness. It reflects the idea that idleness is a cost to be reduced.

        Idleness is not necessarily to be considered solely as a cost. For instance, \cite{russell1932praise}  argued that idleness is required to ensure the preservation of both workers' and machines' capabilities through periods of inactivity. An extensive use of funds can lead to breakdowns, in the form of overheating machines or exhausted workers whose risk of injury is increased. While these considerations are not directly modelled here, they are worth acknowledgement. 
        
        Instead, in this paper---and in line with some elements in \cite{cohendetetal2024}---we consider an alternative way to keep the workers employed despite an increase in intentional idle rates: the generation of innovative ideas. We assume that workers are well-suited for identifying sources of inefficiency and proposing ideas that, when implemented as innovations, could improve execution of their tasks. We assume these process innovations to be incremental and time-saving, hence reducing the duration $T_h$ for specific phases. 
        
        The firm manages the implementation of these ideas through its R\&D department. While idleness-driven innovations might not be as disruptive as those from high-skilled R\&D teams, they should be unlikely to harm the production process since the workers generating them are in direct contact with the production system. Equally important here is the idea that the workers in contact with the production process are less likely to request innovative ideas that could impede the safety of the task.

    \subsection{Modelling of Innovation}  \label{sec:inno}
    
        \textbf{Idea Generation} In direct line with the Schumpeterian, evolutionary literature, the process of innovation is to be understood as a dynamic, uneven but endogenous mechanism of improvement of the technological characteristics of the firm (here the production phases) requiring prior investments in the resources dedicated to the innovation process. These improvements more particularly concern the productivity of a given production phase, while the resource required to generate novelty (ideas here) is time.
        
        Let $t^-_{i, t}$ be the accumulated idle time of worker $i$ at time $t$ and $t_i^*$ takes on the last period at which the worker has had a creative idea and 1 if it has never happened yet. The probability for idleness-driven ideas to be generated by any worker $i$ is an increasing function of their accumulated idle time $\mathcal T_{i,t}^-$, defined as:        
        \begin{equation}
            \mathcal T_{i,t}^- =  \sum_{z = t^*_i}^{t} t^-_{i, z} = 
                \sum_{z = t^*_i}^{t} (1 - f_{i,z})
        \end{equation}

        Workers generate innovative ideas at time $t$ with probability:
        \begin{equation}
            \text P_{i,t} (\text{idea}) = \frac{T_{h_i^*,t}}{g} \left(1 - e^{-\kappa \mathcal T^-_{i,t}} \right)
        \end{equation}            
        
        The probability is increasing in the phase duration $T_{h^*_i,t}$---with $h^*_i$ being the phase the worker $i$ has been hired for---to account for the struggle to improve more efficient tasks. $T_{h^*_i,t}/g$ is the highest probability the worker can reach, and $\kappa$ acts on the idle time necessary to reach the latter asymptote. 

        \hfill\\
        \textbf{Impact} When implemented, a process innovation originating from the worker $i$'s idea has an impact that depends on their productivity level as of the moment $t_i^*$ they produce the idea, i.e., $a_{i,h^*_i,t_i^*}$. Letting $\zeta\in(0,1)$ control the step size of innovations, we define:
        \begin{equation*} 
            T_{h^*_i,t} = \parent{1 - \zeta\cdot \tilde\alpha_{i,h^*_i} } T_{h^*_i, t-1}, \quad \tilde\alpha_{i,h^*_i} \sim\mathcal U\parent{0,\ a_{i,h^*_i,t_i^*}} 
        \end{equation*}
    
        We assume that innovation on phase $h$ is halted as soon as the firm achieves $T_h\leq1$. Moreover, each process innovation implies some degree of disruption and, as a result, workers must adapt to the new features. To model this we update the productivity level of workers on the targeted phase and let the reduction be proportional to both the degree of change and the forgetting threshold:
        \begin{equation}\label{eq:update_wp}
            a_{i, h^*_i, t} = \max\left\{a_0,\ 
                \parent{1 - \frac{\theta_a \abs{\Delta T_{h^*_i,t}}}{T_{h^*_i,t-1}}} a_{i,h^*_i,t}
            \right\}
        \end{equation}

        \hfill\\
        \textbf{Management and Implementation} The firm considers implementing an idea if the worker's productivity for the targeted task exceeds a threshold $\underline a \geq a_s$. Thus, an idea is added to the stack innovative ideas if:
        \begin{equation}\label{eq:review_ideas}
            a_{i,h^*_i,t_i^*} > \underline a
        \end{equation}        

        The stack is a form of memory allowing the firm to reconsider past ideas later on.\footnote{The idea of a non-ephemeral repertoire of ideas has been coined the \textit{creative slack} \citep{cohendet2007playing} and integrated in new theories of the firm \citep{cohendetetal2024}.} However, this memory is cleared up because we assume that innovative ideas are interdependent and tied to the current state of the production process. Incidentally, when an innovation affects some phase, all pending ideas for that same phase are discarded to avoid the adverse effects of conflicting modifications.

        Every $\tau$ periods, if no R\&D activity is ongoing, the firm reviews the stack and selects the next innovation to implement. The selection is based on the idea submitted by the worker with the highest productivity at the time of submission. The chosen innovation is then forwarded to the R\&D department that needs time to develop it. Once developed, the innovation is implemented in the next planning period, allowing the process to be rescheduled and resources reallocated.
        
        The time-to-build $B_{h^*_i,t}$ is increasing in the innovation's impact: the more time-saving the innovation the longer it takes to be developed. Besides, $B_{h^*_i,t}$ is also a function of the phase duration $T_{h^*}$, again due to the struggle to improve efficient technologies. We define:
        \begin{equation}\label{eq:timetobuild}
             B_{h^*,t} = \beta \cdot \frac{\abs{T_{h^*_i,t}-T_{h^*_i,t-1}}}{\parent{T_{h^*_i, t-1}}^2}
         \end{equation}

    \subsection{Configuration}

        Unless stated otherwise, as part of the sensitivity analysis, the parameters related to the innovation process are set as follows:
        
        \begin{itemize}[partopsep=.8ex, itemsep=0pt]
            \item We run 50 simulations per configuration and conduct a Monte Carlo analysis.
            \item The idea generation parameters are set to $g=10000$ and $\kappa=0.002$.
            \item The innovation step size is set to $\zeta=0.1$.
            \item The acceptation threshold is set to $\underline a=0.2$.
            \item The time-to-build parameter is set to $\beta=10000$.
        \end{itemize}

        Variables and parameters are summarized in table \ref{table} in Appendix \ref{sec:appendix}.

    \subsection{Experiment 4: Idleness-driven Innovation} \label{sec:exp4}

        In order to isolate the effect of process innovation, we set $\gamma_a=0$ and $\theta_a=0$---this neutralizing both learning and forgetting. This also implies that innovations do not affect the workers' productivity. We choose to focus this experiment on the effect of both the frequency and the impact of ideas through the setting of $g\in\{1000,10000,100000\}$ and $\zeta\in\{0, 0.01, 0.05, 0.1, 0.5\}$.

        \begin{figure}[ht]
            \centering
            \includegraphics[width=\textwidth]{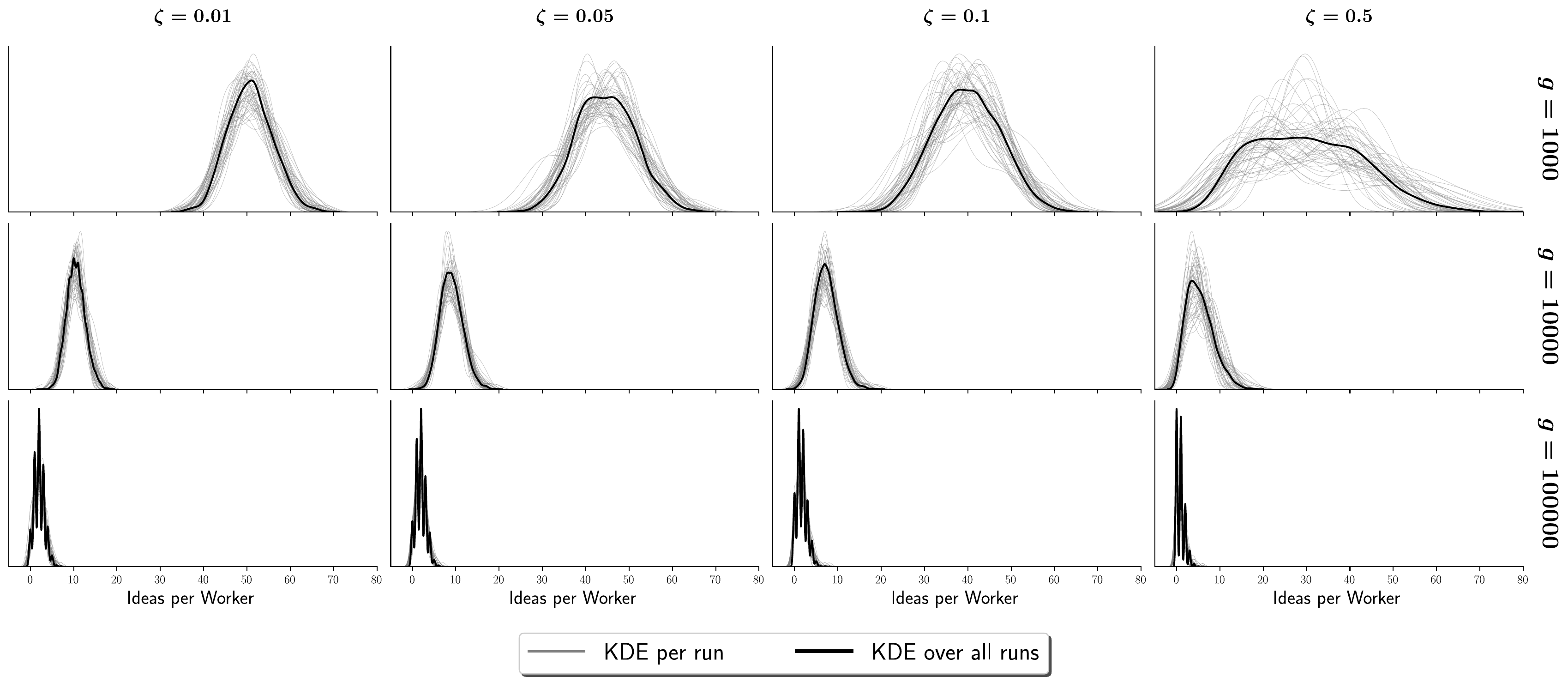}
            \caption{Ideas Distribution across Innovation Regimes. The Kernel Density Estimates (KDEs) are plotted for the 50 Monte Carlo runs. The KDE computed over all runs is shown in bold. The step-size parameter $\zeta$ varies in columns and the inverse frequency of ideas $g$ varies in rows. The scales and the densities are not differentiates so as the keep the eye's attention on the distribution shapes.}
            \label{fig:innoN}
        \end{figure}
        
        Figure \ref{fig:innoN} depicts the distribution of creative productivity (idea generation) across workers.\footnote{Note that ideas do not necessarily lead to an innovation.} Lower values of $g$ imply a higher probability of generating ideas. This mechanically increases the number of ideas per worker. An increase in the innovation step size $\zeta$ has the opposite effect and makes the distribution more uniform. There are at least two reasons for this. On the one hand, a large step size implies that fewer ideas are required to reach the most efficient technology ($T_h\leq1$). On the other hand, as the technology gets more efficient the generation of ideas becomes slower. 

        \begin{figure}[!h]
            \centering
            \includegraphics[width=\textwidth]{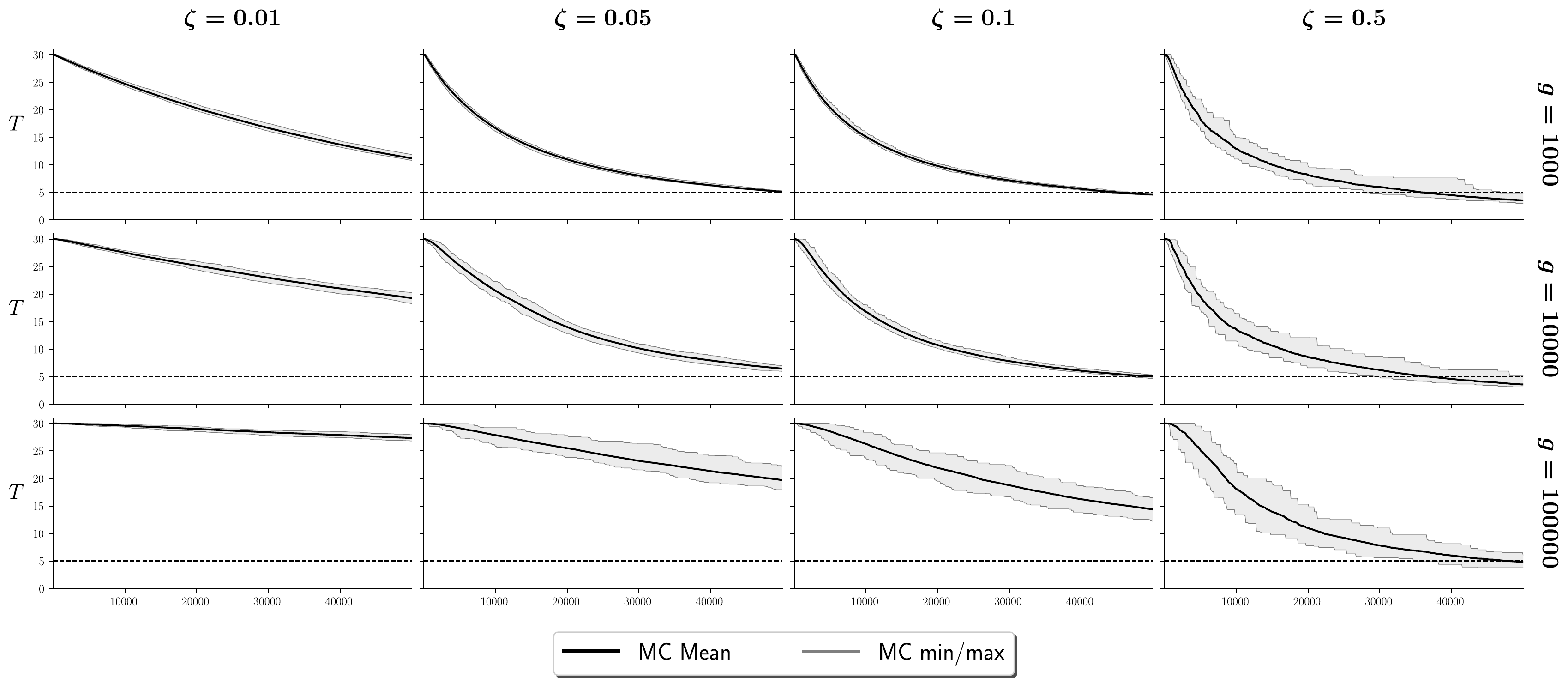}
            \caption{Innovation Pace across Innovation Regimes. Is represented the evolution over time of the total duration $T=\sum_hT_h$ through the Monte Carlo mean, minimum, and maximum values. The step-size parameter $\zeta$ varies in columns and the inverse frequency of ideas $g$ varies in rows. The y-axis is shared across all subplots and a horizontal line is drawn at $T=5$, approximately corresponding to the minimum attainable duration.}
            \label{fig:innoTh}
        \end{figure}

        Figure \ref{fig:innoTh} presents the evolution of the total duration $T$ of the elementary process (under maximum efficiency). Even though both frequency and step size have an expected accelerating effect on the innovation pace, the difference between $g = 1000$ and $g = 10000$ is imperceptible. When the frequency of ideas is high enough, the firm is constantly developing upgrades and thus the ideas pile up too fast. Note, moreover, that a higher step size $\zeta$ is associated with a wider Monte Carlo min-max range. This extra variability arises from the underlying variability and uncertainty of the  impacts of innovation. Furthermore, a higher frequency of ideas, although it does not foster better ideas, is also associated with lower variability. 

        \begin{figure}[ht]
            \centering
            \includegraphics[width=\textwidth]{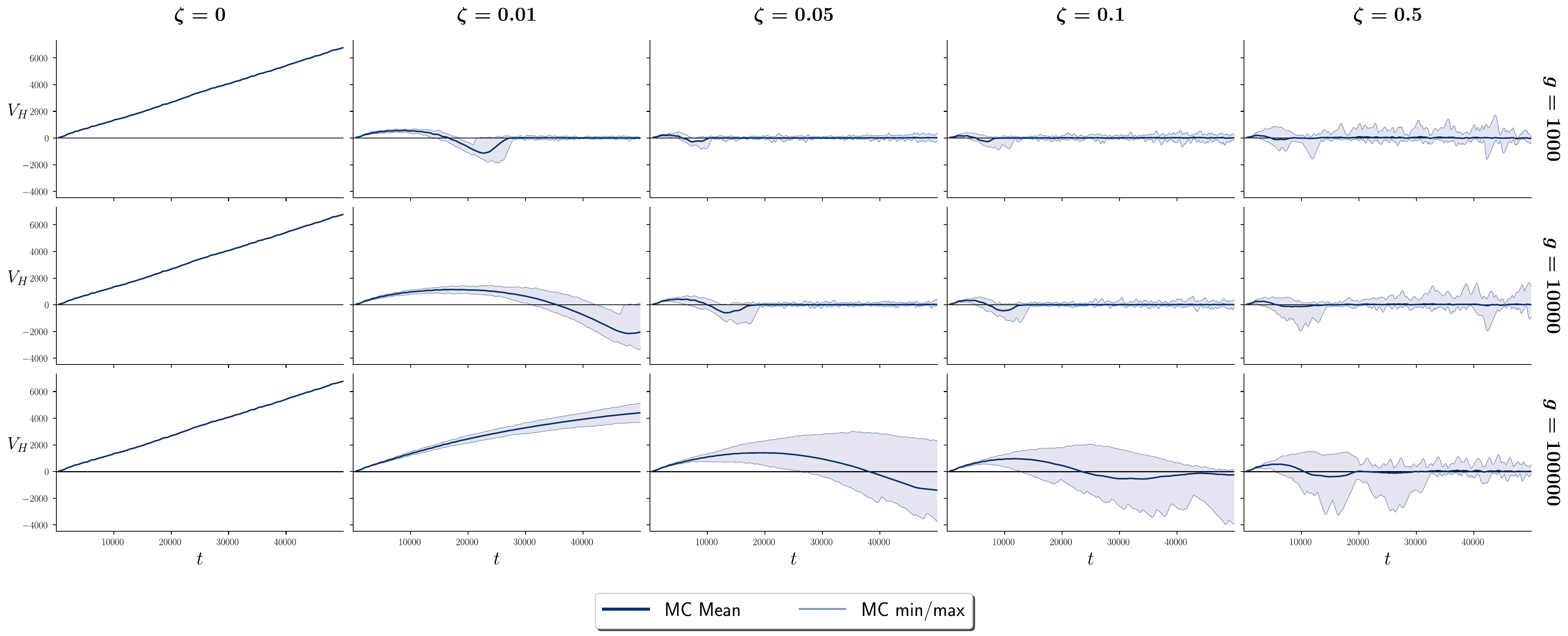}
            \caption{Production Dynamics across Innovation Regimes. The variable $V_H$ smoothed over 100 periods is represented through the Monte Carlo mean, minimum, and maximum values. The step-size parameter $\zeta$ varies in columns and the inverse frequency of ideas $g$ varies in rows. The y-axis is shared across all subplots and a horizontal line is drawn at $V_H=0$.}
            \label{fig:innoVH}
        \end{figure}

        Similar effects on the accumulated production differential $V_H$ are observed (Figure \ref{fig:innoVH}). A larger step size $\zeta$ induces a faster convergence and more long-term variability. A higher frequency of ideas $g$ entails quicker stabilization, up to a certain point. This effect is particularly visible as $\zeta$ increases: a larger step size implies longer development times, so ideas accumulate relatively faster compared to the R\&D capacity. To summarize, more frequent but moderate innovations seem to be the best combination to ensure both a rapid stabilization of the production process and long-term stability.

        \begin{figure}[ht]
            \centering
            \includegraphics[width=\textwidth]{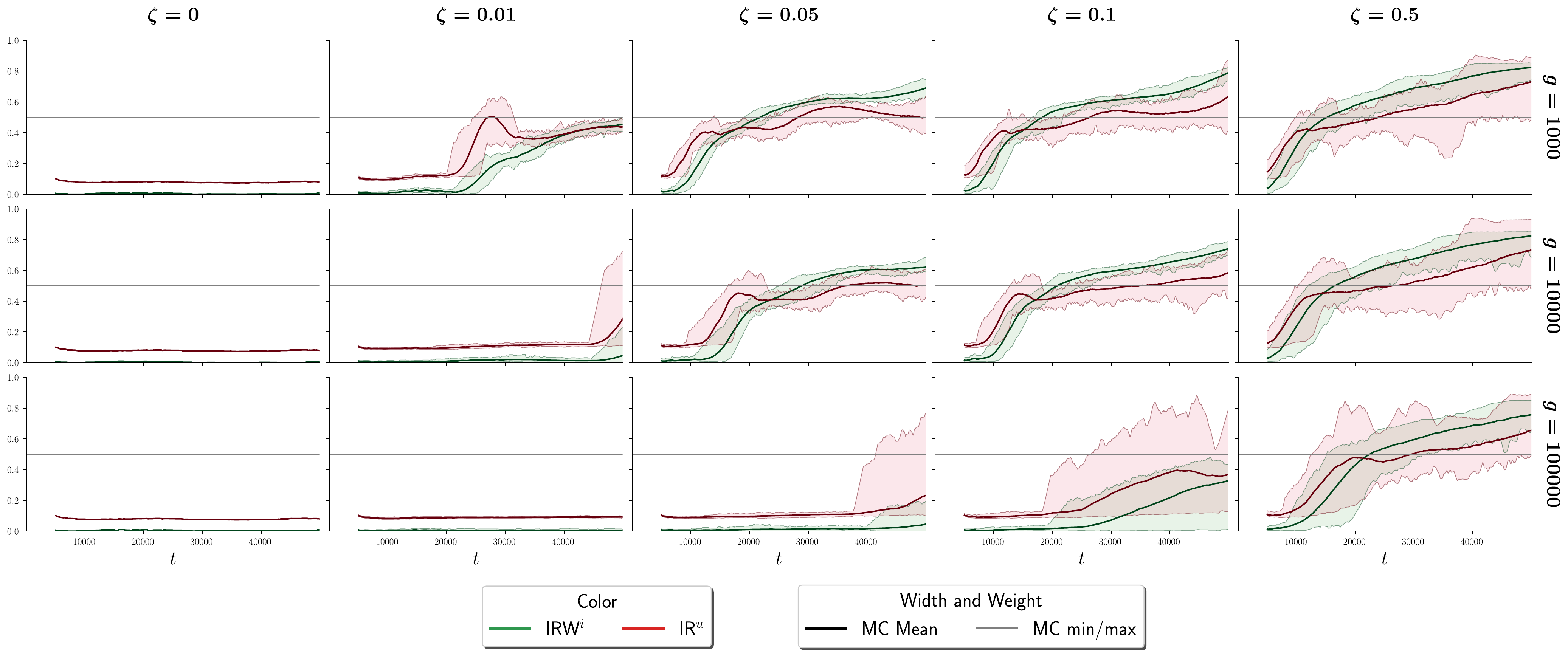}
            \caption{Idleness Rates across Innovation Regimes. The variables IRW$^i$ and IR$^u$ smoothed over 5,000 periods are represented through their Monte Carlo mean, minimum, and maximum values. The step-size parameter $\zeta$ varies in columns and the inverse frequency of ideas $g$ varies in rows. The y-axis is shared across all subplots and a horizontal line is drawn at IR$=0.5$.}
            \label{fig:innoIR}
        \end{figure}

        As shown in Figure \ref{fig:innoIR}, the intentional idleness rate is increasing as the firm innovates. This result is expected as shorter phase durations reduce the need for funds (both workforce and capital). Unintentional idleness also increases as innovations are implemented: both the Monte Carlo mean variability tend to rise over time and are higher when the step size $\zeta$ is large. Moreover, the Monte Carlo variability is higher when ideas are less frequent (high $g$). Because idle rates are strongly smoothed, the Monte Carlo variability mostly depicts the variability across simulations (as opposed to within a simulation), reflecting the diversity of innovation trajectories (strong path dependency). 
        
        This is intertwined with the variability in production discussed earlier. While both intentional and unintentional idleness grow as the firm innovates and the process converges to a stable regime, it is also apparent that the Monte Carlo variability in idleness is connected with the same degree of variability in the production differential. The former phenomenon may result from efficient production leading to periods of excess output, compensated by reactive idleness. The latter, however, is less trivial. The emerging between-run variability in both production and idleness is likely related to shifts in the temporal structures, which, as shown in the previous section, can yield a variety of dynamics. Some trajectories of process duration may indeed be more beneficial to the firm than others, depending on the very temporal structure as well as the MES.

    \subsection{Experiment 5: Innovation and Learning} \label{sec:exp5}

        Now that we have a sense of the effects of idleness-driven innovation, we can explore their interaction with learning dynamics. More formally, we allow the forgetting threshold $\theta_a$ to vary; it controls the differentiation in the learning path of workers, as observed in the dedicated experiment, and affects the impact of innovation on workers' productivity. The learning rate $\gamma_a$ is again set to 0.001. On the idleness-driven innovation side we aim to include both the frequency of ideas and their impact. However, these two parameters are tightly related; for instance, a longer average maturation period for ideas is likely to yield more disruptive process innovations. Therefore, instead of letting both parameters vary independently, we select three scenarios: $(g,\zeta)\in\{(1000,0.01),(10000,0.1),(100000,0.5)\}$, respectively referred to as the 'fast-paced', 'intermediate', and 'disruptive' \textit{creativity policies} (rows 1, 2, and 3 in the graphs). 
        
        Another advantage of treating these parameters as entangled is that they can then be part of the firm's policy toolbox. On the one hand, the fast-paced scenario can be interpreted as a situation in which the firm encourages workers to submit as many incremental ideas as possible. On the other hand, the disruptive scenario would correspond to a firm that advises workers to submit only elaborated, mature ideas. Finally, we differentiate between two managerial approaches: the firm either allows all ideas to accumulate on the stack ($\underline a=0$) or filters out ideas from workers who have started to regress ($\underline a=0.95$), referred to as the 'loose' and 'tight' \textit{stacking policies}, respectively.

        \begin{figure}[ht]
            \centering
            \includegraphics[width=\textwidth]{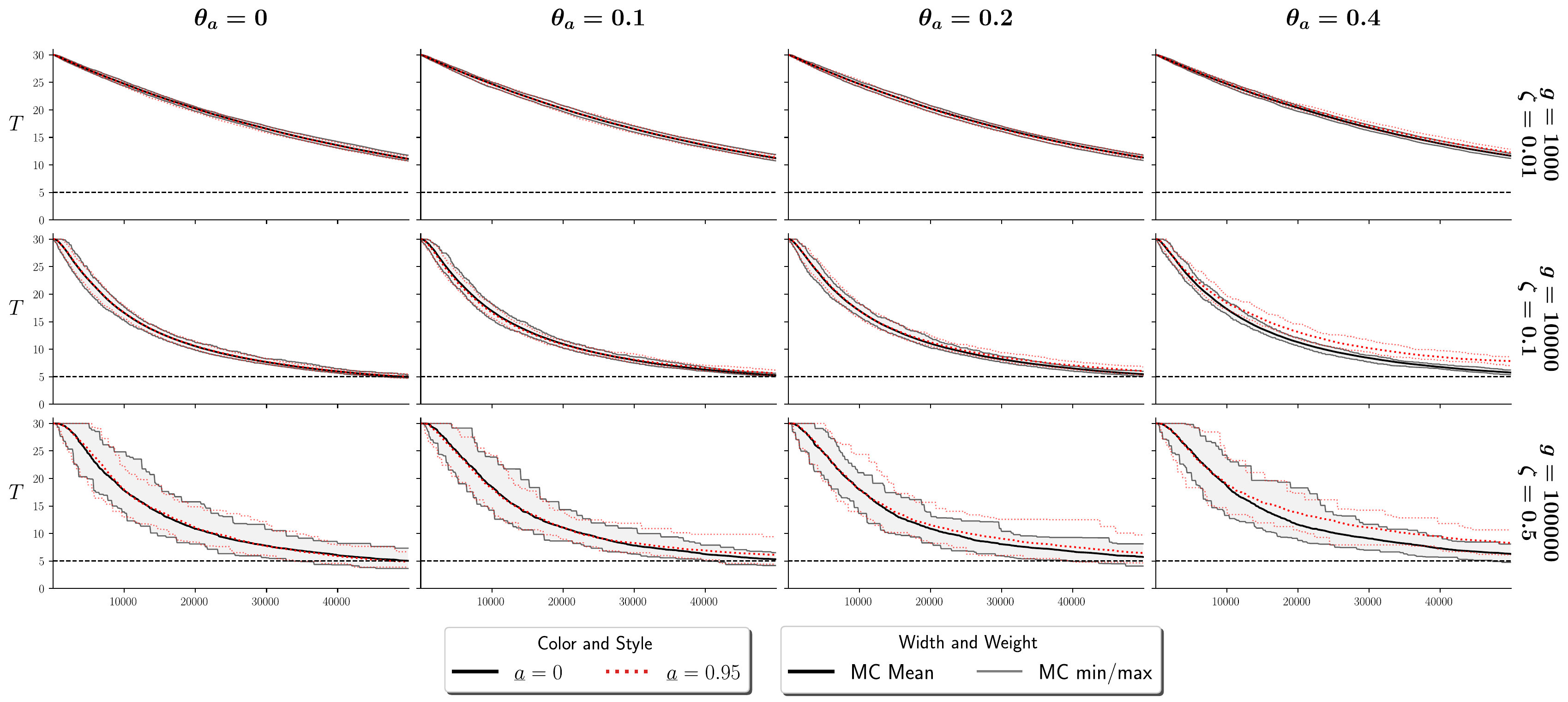}
            \caption{Innovation Pace across Innovation and Forgetting Scenarios. Is represented the evolution over time of the total duration $T=\sum_hT_h$ through the Monte Carlo mean, minimum, and maximum values. The step-size parameter $\zeta$ varies in columns and the inverse frequency of ideas $g$ varies in rows. The two stack policies are differentiate in color and line style. The y-axis is shared across all subplots and a horizontal line is drawn at $T=5$, approximately corresponding to the minimum attainable duration.}
            \label{fig:ILTh}
            
            \includegraphics[width=\textwidth]{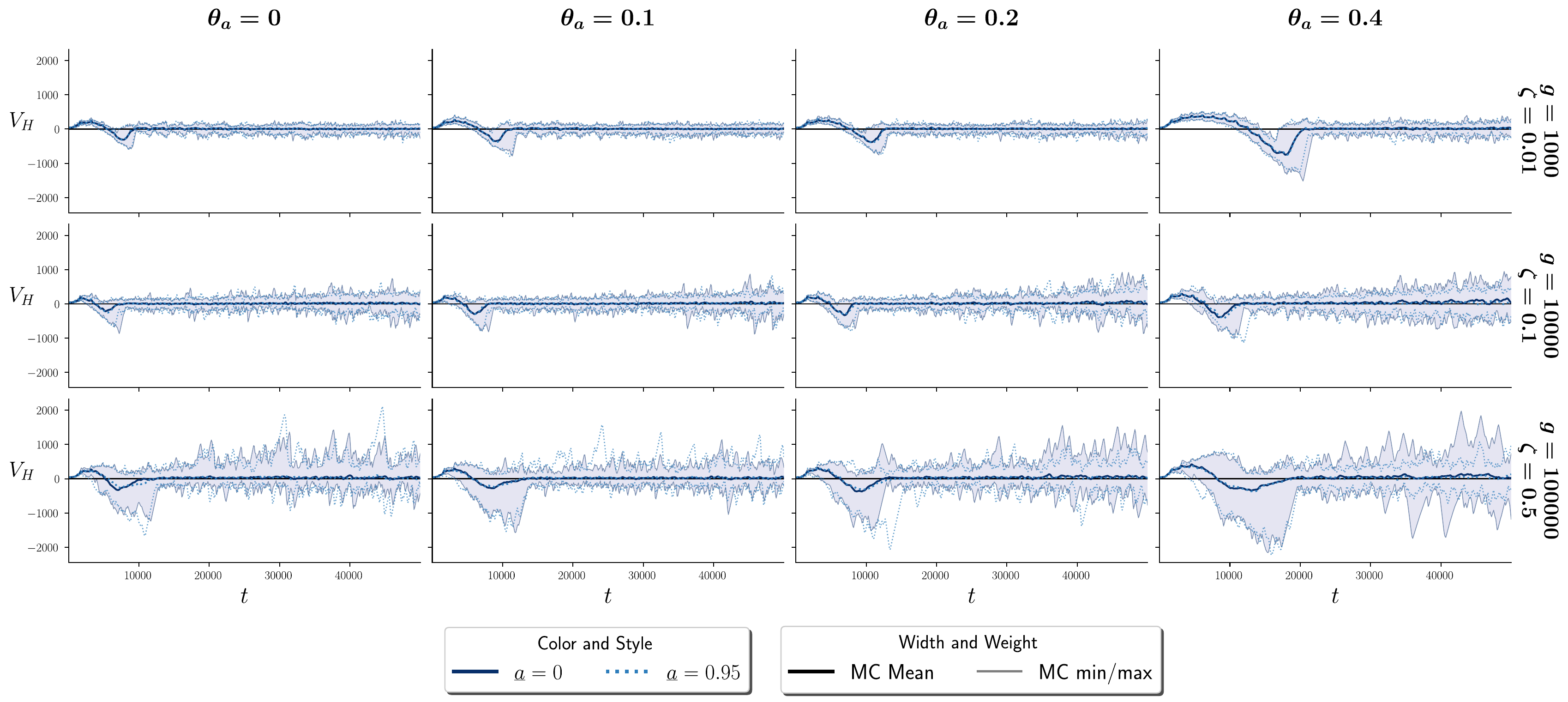}
            \caption{Production Dynamics across Innovation and Forgetting Scenarios. The variable $V_H$ smoothed over 100 periods is represented through the Monte Carlo mean, minimum, and maximum values. The step-size parameter $\zeta$ varies in columns and the inverse frequency of ideas $g$ varies in rows. The two stack policies are differentiate in line style. The y-axis is shared across all subplots and a horizontal line is drawn at $V_H=0$.}
            \label{fig:ILVh}
        \end{figure}

        The effect of the forgetting threshold $\theta_a$ on the pace of innovation is barely noticeable under the loose policy (Figure \ref{fig:ILTh}). However, as the forgetting threshold grows, the gap between the duration dynamics under the two stacking policies widens. Specifically, under the tight policy, innovation visibly slows down as forgetting becomes more pronounced, and this effect is especially noticeable in more disruptive creativity policies. Based on the innovation pace alone, the intermediate scenario seems, nonetheless, the most desirable, as it fosters quick innovation with minimal variability, and both stacking policies appear suitable.

        Similar observations hold for production dynamics. Examining the cumulative production differential $V_H$ in Figure \ref{fig:ILVh}, it is evident that a higher forgetting threshold hinders the convergence to a stable regime and increases the Monte Carlo variability. Variability is also amplified when innovation is more disruptive. Once again, more forgetting and larger innovations highlight the effects of a tight stacking policy. Although the Monte Carlo mean is barely affected, the tight policy dampens variability in these settings. This suggests that, in the presence of more disruptive innovations, the firm should prioritize the most promising ideas and filter out the rest, leading to fewer innovations and, consequently, fewer unnecessary shocks to the process. The impact of the tight policy is even more visible when forgetting is more important as low-tier ideas---responsible for these unnecessary shocks---are more frequent under skill decay.
        
        \begin{figure}[t]
            \centering            
            \includegraphics[width=\textwidth]{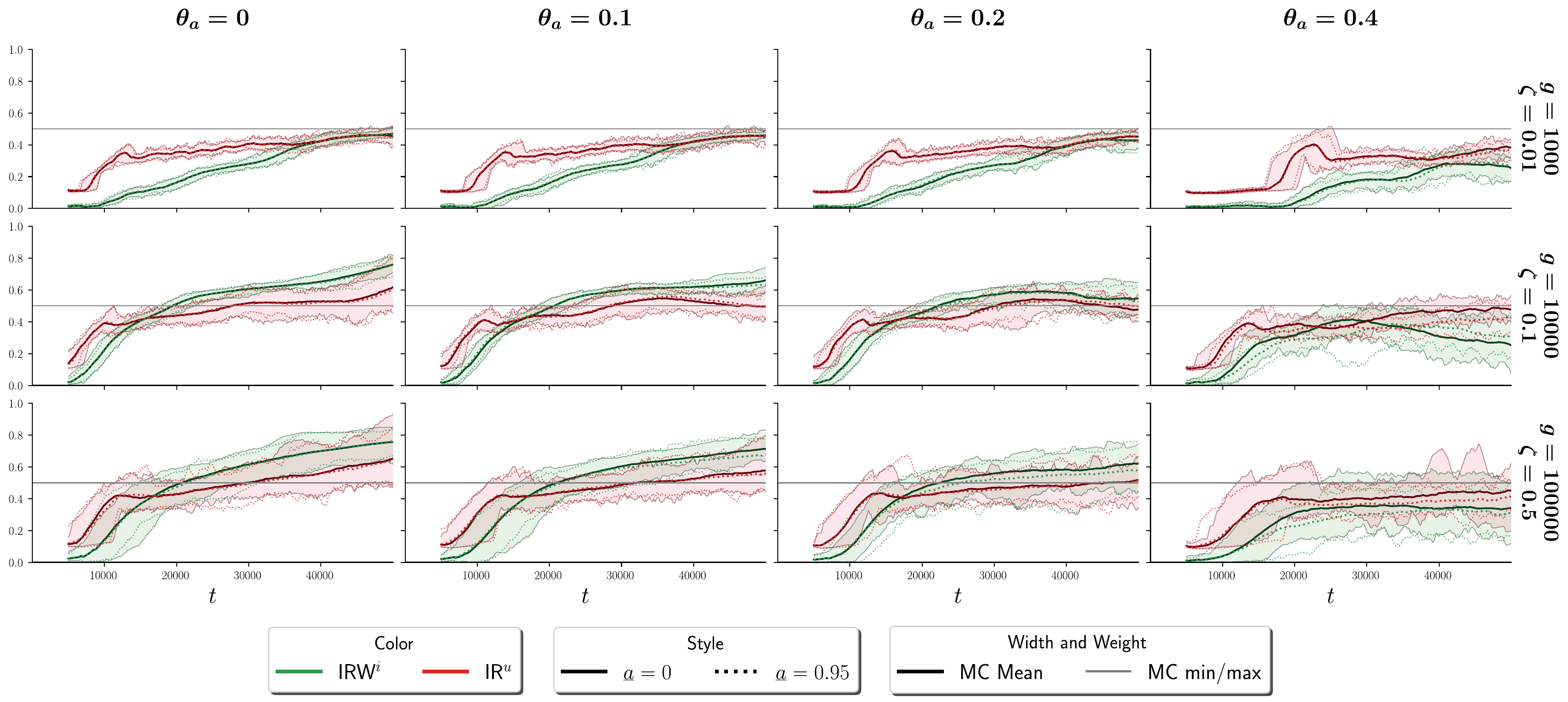}
            \caption{Idleness Rates across Innovation and Forgetting Scenarios. The variables IRW$^i$ and IR$^u$ smoothed over 5,000 periods are represented through their Monte Carlo mean, minimum, and maximum values. The step-size parameter $\zeta$ varies in columns and the inverse frequency of ideas $g$ varies in rows. The two stack policies are differentiate in line style. The y-axis is shared across all subplots and a horizontal line is drawn at IR$=0.5$.}
            \label{fig:ILIR}
            
            \includegraphics[width=\textwidth]{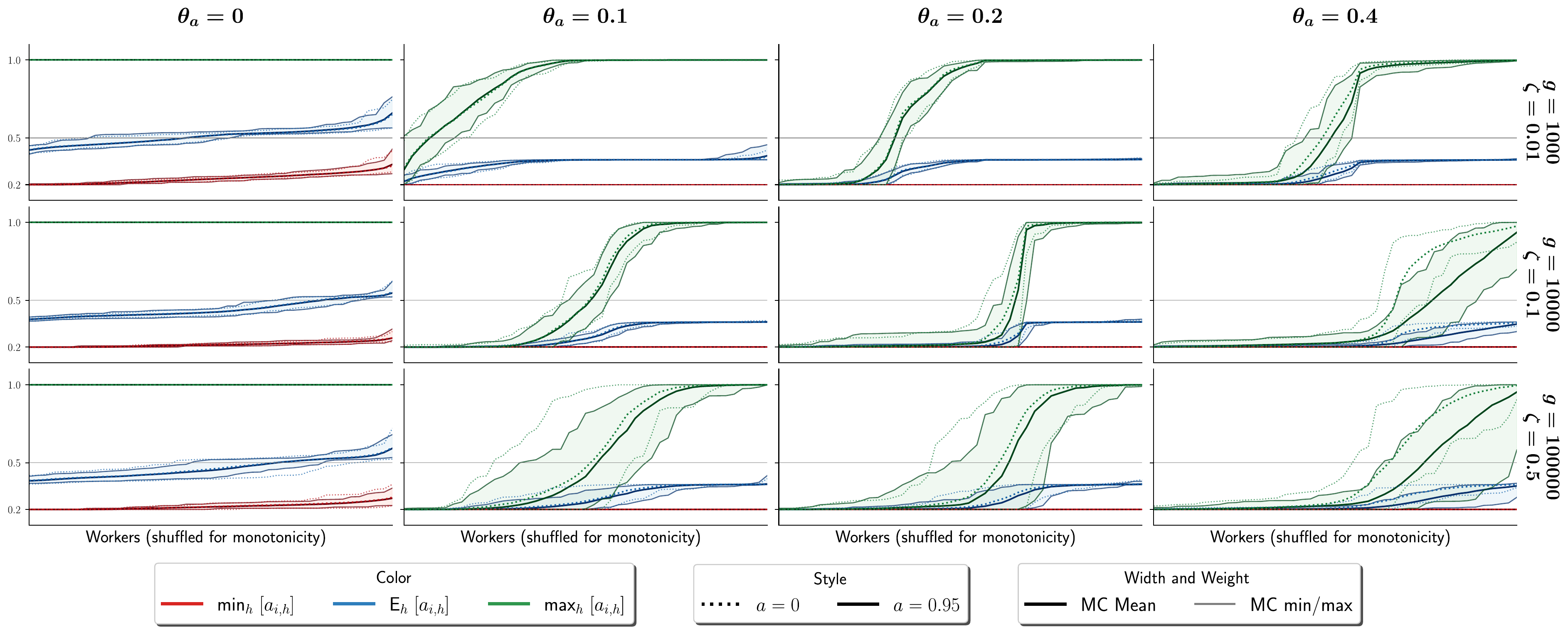}
            \caption{Workers' Specialization across Innovation and Forgetting Scenarios. At the last period we compute the mean, minimum, and maximum of each worker's productivity levels. The resulting series (along workers) are independently sorted and we then graph their Monte Carlo mean, minimum, and maximum values. The graphed series inform on the distribution of productivity levels across workers. The step-size parameter $\zeta$ varies in columns and the inverse frequency of ideas $g$ varies in rows. The two stack policies are differentiate in line style. The y-axis is shared across all subplots and a horizontal line is drawn at $a=0.5$. \textit{Caution: the x-axis is not representing time.}}
            \label{fig:ILPtf}
        \end{figure}

        In the previous section, we observed that forgetting keeps unintentional innovation low in the long run and has little impact on the proportion of workers allocated. Furthermore, innovation without forgetting increased both intentional and unintentional idleness rates. The outcomes with both learning-and-forgetting and innovation dynamics are thus quite surprising. In this context, both intentional and unintentional idleness rates seem to be lower in the long run as the forgetting threshold increases. In Figure \ref{fig:ILIR}, we can even observe cases where average idleness rates decrease over somewhat extended periods of time. A straightforward explanation lies in the effect of forgetting ($\theta_a$) on workers' productivity. Each time an innovation is implemented, workers experience a slight drop in productivity, requiring them to adapt. Consequently, the firm may need to allocate more workers to the upgraded phase shortly after the innovation, even though the duration of this phase is reduced. Moreover, skill decay---due to temporary reinforcements and long-term intentional idleness---offsets the increase in unintentional idleness initially induced by time-saving innovations. This results in more stable idle rates.
        
        These results are especially interesting in that the pace of innovation is sensibly the same across values of $\theta_a$ (Figure \ref{fig:ILTh}). Moreover, we observe a higher Monte Carlo variability in the innovation paces as forgetting becomes stronger. As in the experiment with innovation only, this variability in idleness rates is congruent with the variability in the production differential, tightening up the link between the two. Finally, the effect of a tight stacking policy is positive for the firm and again most visible under the disruptive creativity policy and strong forgetting, wherein long-term idle rates are lower and the additional variability is dampened. 

        Examining the portfolios of skills (Figure \ref{fig:ILPtf}) provides further insight. As expected, and consistent with prior observations, more forgetting leads to more unskilled workers at the end of the simulations. The disruptiveness of innovations has a similar, albeit smaller, effect. Moreover, the same  benefits of a tight policy are visible. Notably, the tight policy results in a more skilled workforce at the end, which aligns with the lower idle rates.

        In a production process wherein workers contribute to process improvements, there is an inherent feedback loop: idleness fosters innovation, and innovations enhance overall productivity, potentially leading to more idleness. More idle workers, however, may experience a drop in productivity, increasing the likelihood of generating ideas of a lower quality. This risk is higher in the most disruptive settings and when forgetting is more pronounced. These results suggest that if the firm stresses the production process by adopting innovations, it should do so at a moderate frequency, focusing on the most promising ideas. This approach would allow workers to adapt as well as reduce uncertainty and instability.

        Figure \ref{fig:ILPtf} further shows that a significant proportion of workers are unskilled by the end of the simulations. This outcome arises for some of the workers who are not allocated for extended periods of time following the adoption of innovations. As this fosters forgetting, they are left out of the process, losing their skills, and thus generate less relevant, outdated ideas. Under a tight stacking policy, these workers are excluded from the innovation process as well, resulting in a high proportion of unskilled workers at the end of the simulation who are indeed useless to the firm.\footnote{This emergent phenomenon of specialisation and asymmetry in the working times is resonating with threshold models of division of labour developed for the study of ants social behaviour \citep{theraulaz1998response}.}

        This reveals an important flaw in the proposed setup and raises questions about the feasibility of the proposed idleness management strategy. To some extent, workers may be \textit{cursed} because, although they remain employed for some time, their contribution to innovation may ultimately turn their service obsolete, both in terms of their productivity on the shop-floor as well as their creative potential. If workers submit many low-tier ideas (loose stacking policy) or fail to produce high-tier ideas (tight stacking policy), the firm in fact has more incentives to dismiss them.

\section{Discussion}

    \textbf{Summary} To reconsider the role of idleness in the context of the co-evolution of production and innovation processes in the firm, we proposed an ABM of the production process, building on the fund-flow approach of NGR that we augment with productivity and innovation dynamics---in line with the Schumpeterian and evolutionary traditions. We further introduce factors indivisibility and let the firm arrange the production system in-line, with the overarching goal of mitigating factors idleness. 
    
    We examined the effects of various organisational choices and observed that indeed the firm's reactivity and proactivity---requiring managerial resources---plays a significant role. Production dynamics converge more rapidly to a stable, low-volatility regime when funds are reallocated frequently, machinery is maintained at a high productivity level, and production objectives are augmented to cope with periods of slowdown (proactivity). These increased degrees of reactivity and pro-activity, however, also lead to higher rates of idleness in the long run. A trade-off therefore emerges between the speed of stabilization and the efficiency of factors utilization.

    Next, we explored the differences in production dynamics across various temporal structures of the sequential production system. A significant degree of heterogeneity has been observed in the resulting dynamics. These highly dissimilar dynamics seem to be explained only to a minor extent by the total duration of the process. Instead, the very temporal structures and their associated MES appeared to have a drastic effect on the stability and convergence of production---with some structures converging rapidly, if not instantly, whereas others getting stuck in a spiral of delay accumulation.  
    
    We also investigated the role of workers' learning dynamics. Their impact on production is straightforward: hiring more proficient workers and having a higher learning rate both positively impact production dynamics, while, on the other hand, the degree of forgetting negatively affects the same dynamics. An emerging result indicates that the diversity of skills---affected by the same parameters---may play a crucial role as well. Phases of production frequently experience slowdowns that can be mitigated by reallocating non-specialist workers so as to reinforce the lagging phases. As swiftly addressing slowdowns is key to performance, it follows that a versatile workforce capable of reinforcing other phases turns out to be an essential asset for the firm. Therefore, specialization should not be viewed as a binary switch but as a spectrum with potential pitfalls at both extremes: while a highly-skilled workforce that retains its expertise is important, it must be balanced with a diversified skill-set.
    
    We then shifted our attention to idleness-driven innovation within the firm, that we present as a two-edged idleness management strategy: it may help (i) leveraging workers' creativity to discover time-saving process innovations and (ii) maintaining employment by avoiding the lay-offs that are (unethically) driven by long-term rises in efficiency and productivity. To isolate the effects of innovation, we first neutralised workers' learning mechanisms. We observed that more frequent but moderately disruptive innovations provide the best combination for ensuring both a fast stabilization of production and long-term stability. Innovation, however, also leads to long-term idleness, as increasing the efficiency of the production process reduces the need for funds, including workers. Additionally, the Monte Carlo variability in both production and idleness dynamics is strongly related to the size of the average innovation jump. More disruptive innovations (more time-saving) result in greater uncertainty in the durations trajectories and thus more heterogeneity in the temporal structures. The latter, in turn, heavily influence the dynamics of production.
    
    Next, we analysed the combined effects of productivity dynamics and idleness-driven innovations, distinguishing between (i) the firm's strategy regarding the acceptance of creative ideas (tight and loose stacking policies) and (ii) three scenarios characterizing the process of idea generation in terms of frequency and impact (fast-paced, intermediate, and disruptive creativity policies). A recurring observation across these experiments is that both stronger forgetting and larger innovation jumps induce greater variability in the dynamics of production, idleness, skills, and durations. In such cases, firms are better off if they adopt a tight stacking policy as it allows to filter out the least promising ideas. This aligns with the hypothesis that larger innovations have more disruptive effects on the production system, requiring workers to adapt to new technologies, affecting the temporal structure, and possibly necessitating a rearrangement of the process. Thus, less frequent but more impactful innovations are preferable, and this is all the more important as the decay in skills is important (because ideas are then less likely to be disruptive).
    
    Interestingly, idleness does not increase as much when productivity dynamics are included alongside innovation. This is likely because innovations reduce workers' productivity temporarily, prompting the firm to allocate more funds (reinforcements) after each innovation. This reduces intentional idleness, whereas unintentional idleness also decreases (or remains stable) due to skill decay. However, this strategy presents a potential drawback: as the process becomes more efficient, some workers are excluded, leading to the decay of their skills and turning their ideas less valuable for the firm. These workers may be sidelined from both the production and innovation processes, or at least, their ideas would only be accepted under a loose stacking policy, albeit at the cost of greater uncertainty and instability for the firm.

    \hfill\\
    \textbf{Future Research} Addressing this issue is essential to ensure that both the workers and the firm benefit from idleness-driven innovation. We propose two tentative solutions. First, introducing a turnover mechanism could maintain regular activity for more workers. For example, instead of consistently assigning the most productive workers first, the firm could periodically assign slightly less productive workers to ensure they consistently practice several tasks. This could foster more stability as reinforcements would be more effective. Second, the firm could dedicate R\&D to product innovation: firms often diversify their products to maintain or increase market share and this often requires the employment of additional funds. In this case, workers could be reassigned to new tasks or apply their skills to a new product line. This approach, combined with idleness-driven process innovations, could benefit the firm while keeping workers engaged in both production and innovation. To fully account for competition and the outcomes of product innovation, these scenarios could be integrated into an explicit market wherein multiple firms compete both in the goods market and in the labour market.
    
    Besides, we identify several avenues for future research building up on the proposed framework. First, economic models are most effective when calibrated to empirical data. With a preliminary understanding of the model’s behaviour, several parameters---particularly those not easily adjustable by the firm---should be set to empirical estimations. For what concerns parameters that are difficult to estimate, such as the forgetting threshold $\theta_a$ and the innovation step size $\zeta$, an algorithmic calibration procedure can be applied \citep{platt2020comparison}. In contrast, parameters such as demand $\mu_d$ and duration $\Th$ could be tailored to specific real-world cases. Relatedly, empirical validation of the model is necessary to assess how well the model reflects \textit{stylized facts}, i.e., how realistic the emergent dynamics are. We must note, however, that micro-economic facts are heterogeneous rather than strongly stylized, hence our decision to abstract from data-driven calibration and validation in this first exploration of the model's dynamics.
    
    Another significant avenue for research involves relaxing the monopolist and constant demand assumptions. Most firms operate in competitive industries where demand dynamics are endogenously determined. Extensions of the model could include (i) the introduction of exogenous demand shocks and more complex stochastic dynamics or (ii) allowing demand to react endogenously to the firm’s ability to meet demand and deliver promptly---mimicking competitive feedbacks without the need to expressly represent the market. Of course, \textit{in fine}, building a competitive market should also be done as, in doing so, one could study heterogeneous markets wherein firms vary in their management strategies and in their initial temporal structures (i.e., their technology). 
    
    Additionally, non-productive tasks such as monitoring, planning, maintenance, or innovation (R\&D) could be modelled as the output of a production process as well. This would allow, on the one hand, to account for outsourcing endeavours alongside capital markets, and on the other hand, to study the role of these tasks in the efficiency of the main production process. Notably, one could question the effect of upstream tasks within a hierarchical organisation on downstream production. In turn, this could shed light on the relative role of the working class, the white-collars, and capitalists in the sustainability and stability of production.
    
    The understanding of the substantial effects of temporal structures on production dynamics is obviously a major gap to be explored. Indeed, the study of temporal structures' effects produced some of the most unexpected results. Identifying the determinants of the speed of convergence to a stable regime could (i) enhance our understanding of the sources of heterogeneity in firm performance and (ii) guide less arbitrary decisions regarding process innovations by identifying the most promising directions in the space of phase durations.

    Regarding innovation, we may want to add depth to the disruptive effect of creative ideas. Although targeted independently, production phases are likely more interdependent than what has been assumed in this paper. And this implies that innovations targeting one phase may affect the neighbouring phases. For example, the execution of one task $h$ may depend on the quality of the execution of the preceding tasks $h-1,h-2,\dots$. This implies that time-saving innovations, although improving some phase $h$, could impede the operation of other tasks. It follows that workers with limited knowledge of the entire process may inadvertently cause negative disruptions. In this context, a diverse skill-set may prove to be more advantageous than mere specialization. While specialization is beneficial for production, broader skills may be more useful for innovation, which ultimately enhances productivity. In these settings, management strategies such as a turnover of the workers may prove helpful in fostering balanced portfolios of skills that ensure both stable production and high-quality process innovations.

    Finally, embedding firms represented by disaggregated processes into macroeconomic models offers another promising direction. For example, supply chains are known for their instability, and micro-founding supply chains could provide insight into how instabilities emerge and propagate upstream and downstream. A key objective could be the identification of firm-level inefficiencies that are the most disruptive and long-lasting within the chain.
    
    Overall, we hope that this model, or refined versions of it, will contribute to a deeper understanding of the inefficiencies of production at both the firm-level and the market-level, and ultimately shed light on how these factors impact our economic societies at large.

\hfill\\
\hfill\\
\hfill\\
\rule{\textwidth}{.05cm}\\ \vspace{.5cm}

\hfill\\
\textbf{Declarations} This work and the publication have been supported by the Chair Management of Creativity, funded by Schmidt Groupe and the Fondation Université de Strasbourg. 
The authors have no competing interests to declare.

\hfill\\
\textbf{Code Availability} 
Simulations have been modeled and run using LSD \citep{pereira2023lsd}, an open free software for ABM development (GNU General Public License, Copyright Marcelo Pereira and Marco Valente). The code of the simulation model and the analyses are available upon request from the referees and the readers. 

\hfill\\
\textbf{Acknowledgments} 
The authors wish to thank Luigi Marengo and Marco Valente for this paper benefited greatly from discussions started more than 10 years ago while working on \cite{LlerenaLorentzMarengoValente}. 
The current paper also owes to the comments received to earlier works presented by A. Lorentz and P. Llerena respectively at the 2014 International Schumpeter Society Conference in Jena and at HEC Montreal in 2018 as well as the 2021 International Schumpeter Society Conference in Rome. The authors are grateful to Laurent Simon and Patrick Cohendet from HEC Montreal (Canada), as well as to Giovanni Dosi and coordinators of the LEM Seminars at Scuola Superiore Sant'Anna di Pisa (Italy), and to the participants of the Autumn School of Creativity Management (CreaSXB) in Strasbourg (2023) for insightful discussions.

\newpage
\bibliographystyle{abbrvnat} 
\bibliography{main}

\appendix

\section{Summary Table}\label{sec:appendix}

\vspace{.5cm}
\begin{adjustwidth}{-2cm}{-2cm}

    \centering
    \setlength{\arrayrulewidth}{.1em}
    \renewcommand{\arraystretch}{1.5}
    \resizebox{1.1\textwidth}{!}{%

    \begin{tabular}{c|llllll}

        \rowcolor[HTML]{C0C0C0} 
          \multicolumn{1}{c|}{\cellcolor[HTML]{C0C0C0}\Large\textbf{Level}} &
          \multicolumn{1}{c}{\cellcolor[HTML]{C0C0C0}\Large\textbf{Type}} &
          \multicolumn{1}{c}{\cellcolor[HTML]{C0C0C0}\Large\textbf{Notation}} &
          \multicolumn{1}{c}{\cellcolor[HTML]{C0C0C0}\Large\textbf{Description}} &
          \multicolumn{1}{c}{\cellcolor[HTML]{C0C0C0}\Large\textbf{Initial Value}} &
          \multicolumn{1}{c}{\cellcolor[HTML]{C0C0C0}\Large\textbf{Definition}} &
          \multicolumn{1}{c}{\cellcolor[HTML]{C0C0C0}\Large\textbf{Analysis}} \rule{0pt}{5ex} \\[2ex]  \hline
        \rowcolor[HTML]{EFEFEF} 
        \cellcolor[HTML]{C0C0C0} &
          \cellcolor[HTML]{EFEFEF} &
          $T, T^*$ &
          Total process duration at maximum efficiency &
          30 &
          \ref{sec:seq}, \ref{sec:plan} &
          \ref{sec:exp2} \\
        \rowcolor[HTML]{EFEFEF} 
        \cellcolor[HTML]{C0C0C0} &
          \cellcolor[HTML]{EFEFEF} &
          $\delta, \delta^*$ &
          Elementary lag (cycle time) of the in-line process &
          6 &
          \ref{sec:seq}, \ref{sec:plan} &
          \ref{sec:exp2} \\
        \rowcolor[HTML]{EFEFEF} 
        \cellcolor[HTML]{C0C0C0} &
          \cellcolor[HTML]{EFEFEF} &
          $n^*$ &
          Number of lines &
          6 &
          \ref{sec:plan} &
           \\
        \rowcolor[HTML]{EFEFEF} 
        \cellcolor[HTML]{C0C0C0} &
          \cellcolor[HTML]{EFEFEF} &
          $C_k, C_k^*$ &
          Minimum efficient size &
          30 &
          \ref{sec:plan} &
           \\
        \rowcolor[HTML]{EFEFEF} 
        \cellcolor[HTML]{C0C0C0} &
          \cellcolor[HTML]{EFEFEF} &
          $V_H$ &
          Cumulative production differential (final product) &
          0 &
          \ref{sec:perf} &
          \ref{sec:exp1}, \ref{sec:exp2}, \ref{sec:exp3}, \ref{sec:exp4}, \ref{sec:exp5} \\
        \rowcolor[HTML]{EFEFEF} 
        \cellcolor[HTML]{C0C0C0} &
          \cellcolor[HTML]{EFEFEF} &
          IRW$^i$ &
          Workers' intentional idle rate &
           &
          \ref{sec:perf} &
          \ref{sec:exp1}, \ref{sec:exp3}, \ref{sec:exp4}, \ref{sec:exp5} \\
        \rowcolor[HTML]{EFEFEF} 
        \cellcolor[HTML]{C0C0C0} &
          \multirow{-7}{*}{\cellcolor[HTML]{EFEFEF}\large\textbf{Variable}} &
          IR$^u$ &
          Unintentional idle rate &
           &
          \ref{sec:perf} &
          \ref{sec:exp1}, \ref{sec:exp3}, \ref{sec:exp4}, \ref{sec:exp5} \\
        \cellcolor[HTML]{C0C0C0} &
           &
          $\mu_d$ &
          Demand &
          1 &
          \ref{sec:plan} &
           \\
        \cellcolor[HTML]{C0C0C0} &
           &
          $\tau$ &
          Planning periodicity &
          50 &
          \ref{sec:plan} &
          \ref{sec:exp1} \\
        \cellcolor[HTML]{C0C0C0} &
           &
          $r$ &
          Proactivity multiplier &
          1.5 &
          \ref{sec:fund} &
          \ref{sec:exp1} \\
        \cellcolor[HTML]{C0C0C0} &
           &
          $H$ &
          Number of phases &
          5 &
          \ref{sec:seq} &
           \\
        \cellcolor[HTML]{C0C0C0} &
           &
          $\zeta$ &
          Innovation step size &
          0.1 &
          \ref{sec:inno} &
          \ref{sec:exp4}, \ref{sec:exp5} \\
        \cellcolor[HTML]{C0C0C0} &
           &
          $\underline a$ &
          Stacking policy (ideas screening) &
          0.2 &
          \ref{sec:inno} &
          \ref{sec:exp5} \\
        \multirow{-14}{*}{\cellcolor[HTML]{C0C0C0}\Large\textbf{Firm}} &
          \multirow{-7}{*}{\large\textbf{Parameter}} &
          $\beta$ &
          Time-to-build parameter &
          10000 &
          \ref{sec:inno} &
           \\ \hline
        \rowcolor[HTML]{EFEFEF} 
        \cellcolor[HTML]{C0C0C0} &
          \cellcolor[HTML]{EFEFEF} &
          $T_h$ &
          Phase duration &
          6 &
          \ref{sec:seq}, \ref{sec:plan}, \ref{sec:inno} &
          \ref{sec:exp2}, \ref{sec:exp4}, \ref{sec:exp5} \\
        \rowcolor[HTML]{EFEFEF} 
        \cellcolor[HTML]{C0C0C0} &
          \cellcolor[HTML]{EFEFEF} &
          $I_h$ &
          Phase inflows stock &
          0 &
          \ref{sec:prod} &
           \\
        \rowcolor[HTML]{EFEFEF} 
        \cellcolor[HTML]{C0C0C0} &
          \cellcolor[HTML]{EFEFEF} &
          $Q_h$ &
          Phase production &
           &
          \ref{sec:prod} &
           \\
        \rowcolor[HTML]{EFEFEF} 
        \cellcolor[HTML]{C0C0C0} &
          \cellcolor[HTML]{EFEFEF} &
          $C_h, C_h^*$ &
          Efficient number of duos &
          6 &
          \ref{sec:plan} &
           \\
        \rowcolor[HTML]{EFEFEF} 
        \cellcolor[HTML]{C0C0C0} &
          \cellcolor[HTML]{EFEFEF} &
          $e_h$ &
          Maintenance cost &
          500/6 &
          \ref{sec:fund} &
           \\
        \rowcolor[HTML]{EFEFEF} 
        \cellcolor[HTML]{C0C0C0} &
          \cellcolor[HTML]{EFEFEF} &
          $M_h$ &
          Number of allocated duos &
           &
          \ref{sec:fund} &
           \\
        \rowcolor[HTML]{EFEFEF} 
        \cellcolor[HTML]{C0C0C0} &
          \cellcolor[HTML]{EFEFEF} &
          $J_h$ &
          Initial number of machines &
          12 &
          \ref{sec:fund} &
           \\
        \rowcolor[HTML]{EFEFEF} 
        \cellcolor[HTML]{C0C0C0} &
          \cellcolor[HTML]{EFEFEF} &
          $N_h$ &
          Initial number of workers &
          9 &
          \ref{sec:fund} &
           \\
        \rowcolor[HTML]{EFEFEF} 
        \multirow{-9}{*}{\cellcolor[HTML]{C0C0C0}\shortstack{\Large\textbf{Production}\\[2ex]\Large\textbf{Phase}}} &
          \multirow{-9}{*}{\cellcolor[HTML]{EFEFEF}\large\textbf{Variable}} &
          $V_h$ &
          Cumulative production differential &
          0 &
          \ref{sec:perf} &
           \\ \hline
        \cellcolor[HTML]{C0C0C0} &
           &
          $f_{i/j,h}$ &
          Fractional working time of fund $i$ or $j$ on phase $h$ &
           &
          \ref{sec:fund} &
           \\
        \cellcolor[HTML]{C0C0C0} &
           &
          $q_{ij,h}$ &
          Latent production by the duo $ij$ on phase $h$ &
          0 &
          \ref{sec:fund} &
           \\
        \cellcolor[HTML]{C0C0C0} &
           &
          $a_{i,h}$ &
          Worker productivity on phase $h$ &
          $a_s$ on $h_i^*$, else $a_u$ &
          \ref{sec:fund} &
          \ref{sec:exp3}, \ref{sec:exp5} \\
        \cellcolor[HTML]{C0C0C0} &
           &
          $\mathcal T^-_i$ &
          Accumulated worker's idle time &
          0 &
          \ref{sec:inno} &
           \\
        \cellcolor[HTML]{C0C0C0} &
           &
          $\text P_i$ &
          Worker's idea generation probability &
          0 &
          \ref{sec:inno} &
           \\
        \cellcolor[HTML]{C0C0C0} &
           &
          $B_h$ &
          Time-to-build &
           &
          \ref{sec:inno} &
           \\
        \cellcolor[HTML]{C0C0C0} &
           &
          $F_{j,h}$ &
          Accumulated machine's working time since repair &
           &
          \ref{sec:fund} &
           \\
        \cellcolor[HTML]{C0C0C0} &
          \multirow{-8}{*}{\large\textbf{Variable}} &
          $b_{j,h}$ &
          Machine productivity &
          1 &
          \ref{sec:fund} &
           \\ 
        \rowcolor[HTML]{EFEFEF} 
        \cellcolor[HTML]{C0C0C0} &
          \cellcolor[HTML]{EFEFEF} &
          $h_i^*$ &
          Worker's specialty &
           &
          \ref{sec:fund} &
           \\
        \rowcolor[HTML]{EFEFEF} 
        \cellcolor[HTML]{C0C0C0} &
          \cellcolor[HTML]{EFEFEF} &
          $a_s$ &
          Hiring productivity &
          1 &
          \ref{sec:fund} &
          \ref{sec:exp3} \\
        \rowcolor[HTML]{EFEFEF} 
        \cellcolor[HTML]{C0C0C0} &
          \cellcolor[HTML]{EFEFEF} &
          $a_u$ &
          Minimum worker productivity &
          0.2 &
          \ref{sec:fund} &
           \\
        \rowcolor[HTML]{EFEFEF} 
        \cellcolor[HTML]{C0C0C0} &
          \cellcolor[HTML]{EFEFEF} &
          $\gamma_a$ &
          Learning-and-forgetting rate &
          0.001 &
          \ref{sec:fund} &
          \ref{sec:exp3} \\
        \rowcolor[HTML]{EFEFEF} 
        \cellcolor[HTML]{C0C0C0} &
          \cellcolor[HTML]{EFEFEF} &
          $\theta_a$ &
          Forgetting degree &
          0.2 &
          \ref{sec:fund}, \ref{sec:inno} &
          \ref{sec:exp3} \\
        \rowcolor[HTML]{EFEFEF} 
        \cellcolor[HTML]{C0C0C0} &
          \cellcolor[HTML]{EFEFEF} &
          $\theta_b$ &
          Depreciation rate &
          0.0002 &
          \ref{sec:fund} &
           \\
        \rowcolor[HTML]{EFEFEF} 
        \cellcolor[HTML]{C0C0C0} &
          \cellcolor[HTML]{EFEFEF} &
          $\underline b$ &
          Maintenance threshold &
          0.8 &
          \ref{sec:fund} &
          \ref{sec:exp1} \\
        \rowcolor[HTML]{EFEFEF} 
        \cellcolor[HTML]{C0C0C0} &
          \cellcolor[HTML]{EFEFEF} &
          $\omega$ &
          Maintenance cost parameter &
          10 &
          \ref{sec:fund} &
           \\
        \rowcolor[HTML]{EFEFEF} 
        \cellcolor[HTML]{C0C0C0} &
          \cellcolor[HTML]{EFEFEF} &
          $g$ &
          Idea generation frequency parameter &
          10000 &
          \ref{sec:inno} &
          \ref{sec:exp4}, \ref{sec:exp5} \\
        \rowcolor[HTML]{EFEFEF} 
        \multirow{-18}{*}{\cellcolor[HTML]{C0C0C0}\Large\textbf{Fund}} &
          \multirow{-10}{*}{\cellcolor[HTML]{EFEFEF}\large\textbf{Parameter}} &
          $\kappa$ &
          Idea generation probability growth rate &
          0.002 &
          \ref{sec:inno} &
          \ref{sec:exp4}
          
    \end{tabular}%
    }

    \label{table}
    \captionof{table}{Summary table}

\end{adjustwidth}

\end{document}